\documentclass[10pt,conference]{IEEEtran}
\usepackage{cite}
\usepackage{amsmath,amssymb,amsfonts}
\usepackage{algorithmic}
\usepackage{graphicx}
\usepackage{textcomp}
\usepackage{xcolor}
\usepackage{multirow}
\usepackage{diagbox}
\usepackage{bm}
\usepackage{epsfig}
 \usepackage{booktabs}
\usepackage{tikz}
\usepackage{changes}
\usepackage{hyperref}
\hypersetup{
    colorlinks=true,       
    linkcolor=blue,        
    citecolor=blue,        
    urlcolor=blue          
}
\def\BibTeX{{\rm B\kern-.05em{\sc i\kern-.025em b}\kern-.08em
    T\kern-.1667em\lower.7ex\hbox{E}\kern-.125emX}}

\title{Top-Down Microarchitecture Analysis Without Causal Modelling Is Dangerous!} 
\author{%
  \IEEEauthorblockN{Chenxi Wang\textsuperscript{1,3}, Lei Wang\textsuperscript{1,2,3}, Wanling Gao\textsuperscript{1,2,3}, Guoxin Kang\textsuperscript{1,2},
  Fanda Fan\textsuperscript{1,2}, Jianfeng Zhan\textsuperscript{1,2}}
  \IEEEauthorblockA{\textsuperscript{1}Institute of Computing Technology, Chinese Academy of Sciences, Beijing, China\\
    \textsuperscript{2}The International Open Benchmark Council, Hong Kong, China\\
    \textsuperscript{3}University of Chinese Academy of Sciences, Beijing, China\\
    Email: \{wangchenxi21s, wanglei\_2011, gaowanling, kangguoxin, fanfanda, zhanjianfeng\}@ict.ac.cn}\\    
}

\begin{document}
\begin{titlepage} 

	\centering 
	
	\scshape 
	
	\vspace*{\baselineskip} 
	
	
	\rule{\textwidth}{1.6pt}\vspace*{-\baselineskip}\vspace*{2pt} 
	\rule{\textwidth}{0.4pt} 
	
	\vspace{0.75\baselineskip} 
	
	{\LARGE Top-Down Microarchitecture Analysis Without Causal Modelling Is Dangerous!} 
	
	\vspace{0.75\baselineskip} 
	
	\rule{\textwidth}{0.4pt}\vspace*{-\baselineskip}\vspace{3.2pt} 
	\rule{\textwidth}{1.6pt} 
	
	\vspace{2\baselineskip} 
	
	
	
	\vspace*{3\baselineskip} 
	
	
	Edited By
	
	\vspace{0.5\baselineskip} 
	
	{\scshape\Large Chenxi Wang\\ Lei Wang\\ Wanling Gao\\ Guoxin Kang\\ Fanda Fan\\ Jianfeng Zhan\\}

	\vspace{0.5\baselineskip} 

	\vfill 
	
	
	\epsfig{file=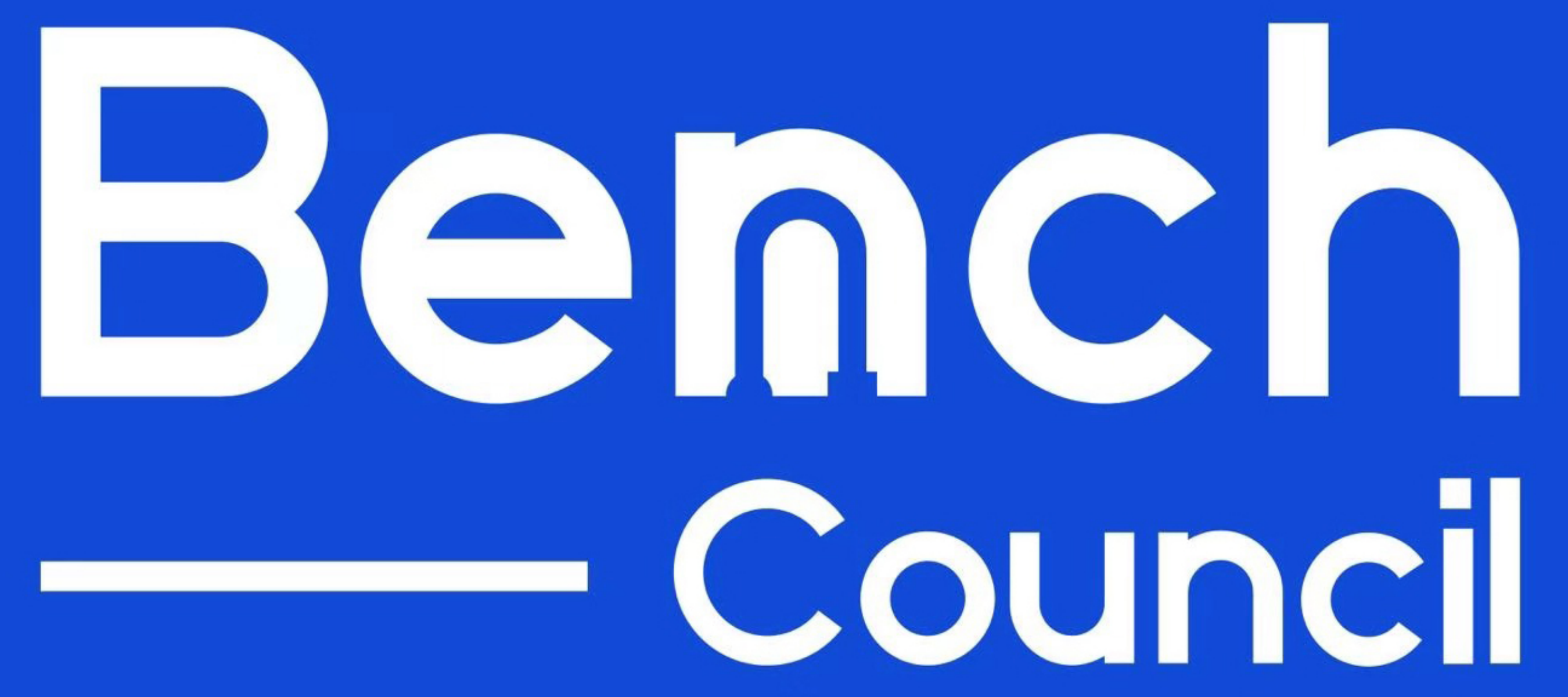,height=2cm}
	\textit{\\BenchCouncil: International Open Benchmark Council\\http://www.benchcouncil.org} 
	\vspace{5\baselineskip} 

	Technical Report No. BenchCouncil-CTMA-2026 
	
	{\large Oct 15, 2026} 

\end{titlepage}
\maketitle
\thispagestyle{plain}
\pagestyle{plain}

\begin{abstract}

Top-Down Microarchitecture Analysis (TMA) is the standard for identifying CPU bottlenecks. However, TMA implicitly models performance via a single causal path: $\text{Workload(dataset, algorithm)} \Rightarrow \text{CPU(CPI)} \Rightarrow \text{System(Performance)}$ (microarchitectural efficiency path). It overlooks two hidden causal paths: $\text{Workload(dataset, algorithm)} \Rightarrow \text{Instructions(Retired)} \Rightarrow \text{System(Performance)}$ (algorithmic efficiency path) and $\text{Workload(dataset, algorithm)} \Rightarrow \text{Instructions(Retired)} \Rightarrow \text{CPU(CPI)} \Rightarrow \text{System(Performance)}$ (microarchitecture-algorithm efficiency path). This causal blindness often misleads architects into futile hardware optimizations. To address this, we propose \textbf{Causal Top-Down Microarchitecture Analysis (CTMA)}. Utilizing the object causal model, CTMA explicitly models these three causal paths among the workload, instructions, CPU, and system.

We validate CTMA across Intel, AMD, and ARM platforms. Our experiments reveal counterintuitive insights: 
(1) \textit{CPI Profiling Blindness}: Algorithmic instruction bloat can double elapsed time while the CPI profile remains frozen. 
(2) \textit{Over-Speculation Congestion}: Throttling the Frontend paradoxically improves performance in pointer-chasing algorithms. 
(3) \textit{ILP Window Collapse}: Extreme data topologies shift bottlenecks from Backend ROB capacity to Frontend fill-rate starvation. 
(4) \textit{Op-Cache SMT Overlap}: Ultra-small footprints yield SMT speedups on Intel but cause severe thrashing on AMD. 
(5) \textit{``Ghost State'' Artifact}: Stale pipeline mitigations on Intel Xeon Gold 5120T artificially boost performance by 22.75\% on SPEC CPU2017 and 42.2\% on SPEC CPU2026. Based on this, we propose a dynamic speculation throttling mechanism.

\end{abstract}


\section{Introduction}\label{sec:introduction}

Modern processor design and optimization heavily rely on Top-Down Microarchitecture Analysis (TMA). By categorizing CPU  pipeline slots into 
Frontend Bound, Backend Bound, Retiring, and Bad Speculation, TMA provides a clear snapshot of pipeline bottlenecks and guides pipeline efficiency optimizations (i.e., minimizing CPI). 
Viewed through the lens of causal modeling, TMA implicitly models performance via a single, oversimplified causal path: $\textbf{Workload(dataset, algorithm)} \Rightarrow \textbf{CPU(CPI)} \Rightarrow \textbf{System(Performance)}$. 
In this article, we use the object causal model (OCM) in Evaluatologly as the causal modeling language and methodology~\cite {zhan2025evaluatology,evaluatology-2026-e}.
Here, the causal path $\textbf{A(p)} \Rightarrow \textbf{B(q)}$ denotes that object $A$ (via its property (observed metric) or mechanism $p$) causally impacts object $B$ (altering its property or mechanism $q$).


By fixing the workload inputs, traditional TMA tacitly assume the workload emits a static stream of instructions, placing the root cause of performance variations squarely on the CPU's microarchitectural efficiency (e.g., CPI).
This oversimplified causal path creates severe ``causal blindness,'' completely ignoring two hidden but critical causal paths.

Real-world workloads are algorithmically dynamic: as the topology, distribution, or locality of input dataset changes, algorithms adapt, causing pipeline bottlenecks to drift significantly. 
This activates two completely hidden causal paths. First, the \textit{Algorithmic Efficiency Path}: $\textbf{Workload(dataset, algorithm)} \Rightarrow \textbf{Instructions(Retired)} \Rightarrow \textbf{System(Performance)}$, where algorithmic instructions bloat can silently drastically alter elapsed time while the CPU's CPI remains perfectly stable (Details in Section~\ref{sec:mot:541}). Second, the \textit{Microarchitecture-Algorithm Efficiency Path}: $\textbf{Workload(dataset, algorithm)} \Rightarrow \textbf{Instructions(Retired)} \Rightarrow \textbf{CPU(CPI)} \Rightarrow \textbf{System(Performance)}$, where changes in instructions volume can fundamentally reshape microarchitectural states, artificially altering the CPI (Details in Section~\ref{sec:mot:523}). Ignoring these paths would trick architects into optimizing hardware for what are essentially software anomalies.

\begin{figure*}[h]
\centering
\includegraphics[scale=0.932]{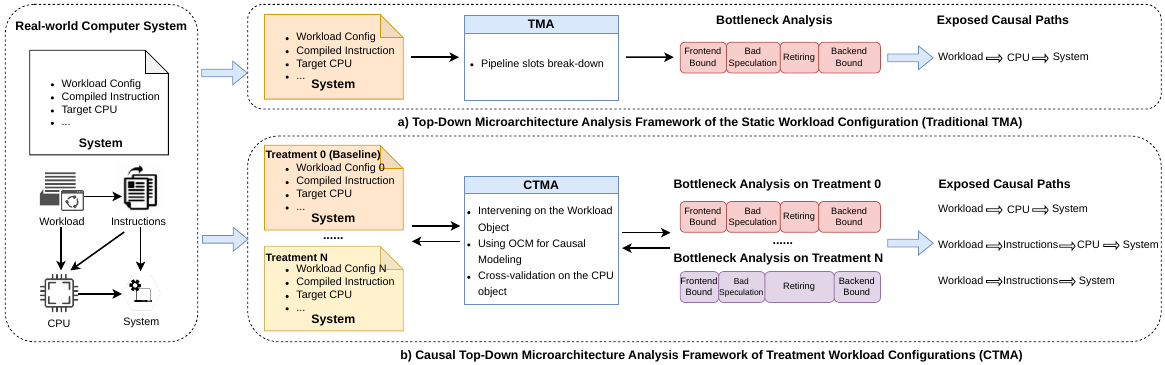}
\caption{
The comparison of traditional TMA and our causal TMA (CTMA).}

\label{fig1-com}
\end{figure*}


To address this critical flaw, we introduce a 
\textbf{Causal Top-Down Microarchitecture Analysis (CTMA)}.  CTMA aims to infer unconfounded architecture insights that can be extended to real-world settings.
In the rest of this article, we use the term \textbf{treatment} to describe the change of a target object while keeping the others the same to understand its causal effect. 

As shown in Figure~\ref{fig1-com}, instead of passively profiling the static workload configuration, CTMA intentionally generates and profiles several workload configuration treatments by mutating specific algorithmic traits (e.g., graph randomness, recursion depth, spatial locality) to expose the TMA-overlooked Algorithmic Efficiency Path and Microarchitecture-Algorithm Efficiency Path. This treatment is to model how the algorithm adapt, and how the pipeline bottlenecks transit shift as the topology, distribution, or locality of input dataset changes. 

Meanwhile, CTMA uses non-invasive hardware treatments to establish unconfounded baselines on real silicon across three major microarchitectural platforms (Intel, AMD, ARM). These hardware platforms feature diverse hardware mechanisms—such as SMT resource partitioning (e.g., partitioning the Reorder Buffer via co-running pause process on the sibling logical core), and core frequency scaling. With the cross-validation of hardware treatments and/or workload configuration treatments, 
CTMA validates all exposed causal paths.







By exposing and controlling the aforementioned three causal paths, CTMA can infer unconfounded architecture insights that can be applied to real-world settings. Utilizing CTMA, we have achieved five counter-intuitive architectural insights.

\begin{itemize}
\item CPI Profiling Blindness (the leela\_r workload from SPEC CPU2017): We prove the existence of ``Algorithmic Instructions Bloat,'' where modifying spatial locality of the dataset causes a more than $100\%$ elapsed time variation while leaving the CPI profile completely frozen. Cross-ISA validation confirmed a universal mathematical constant of $2.06\times$ instructions inflation, effectively proving that microarchitectural tuning is futile against algorithmic degradation.
\item Over-Speculation Congestion (mcf\_r): In pointer-chasing algorithms, highly regular data improves branch prediction but halves performance. Aggressive speculation continuously pushes memory requests into the Backend, overflowing the Line Fill Buffers and causing memory controller congestion. Manually throttling the pipeline by injecting NOP instructions alleviates this traffic jam and slightly improves performance.
\item ILP Window Collapse (cactuBSSN\_r): Spatial topology distortion causes dense short-loops and frequent branch mispredictions. By using SMT to partition the Reorder Buffer (ROB) via pause process, we found that the perfect-topology datasets suffer severe degradation ($\sim$23\% on the Intel Platform and $\sim$12\% on the AMD Platform), while distorted datasets are relatively immune ($\sim$9\%  on the Intel Platform and $\sim$10\% on the AMD Platform). This indicates that extreme dataset distortion shifts the bottleneck from ROB capacity to dynamic fill-rate, rendering deep out-of-order Backend useless.
\item Op-Cache SMT Overlap (xalancbmk\_r): Deep recursion causes severe Frontend Bandwidth bottlenecks. Traditionally, co-running two identical copies of the workload on the two SMT sibling logical cores worsens Frontend contention. However, when the recursion's instructions footprint is extremely small, it resides perfectly within Intel's Decoded Stream Buffer (DSB), enabling cycle-level overlapping and yielding SMT speedups. In contrast, AMD Zen3 suffers more than $50\%$ degradation, exposing rigid Op-cache sharing policies.
\item ``Ghost State'' artifact on the Intel Xeon Gold 5120T: We discovered that forcefully terminating (via \texttt{kill -9}) an SMT sibling process without an OS-level microarchitectural flush leaves the CPU pipeline in a highly aggressive state, resulting in an average of 22.75\% artificial performance boost on SPEC CPU2017 and 42.20\% performance boost on SPEC CPU2026. This exposes a critical measurement artifact linked to spectre\_v2 mitigation toggles on specific unbalanced-designed CPUs. We propose a dynamic speculation throttling mechanism for real-time congestion control in the next-generation CPU.
\end{itemize}

Our contributions are summarized as follows.

\begin{enumerate}

\item For the first time, we introduce the causal modeling methodology, specifically, the object causal model in architecture research.  
\item We expose that TMA based on static workload evaluation severs causal blindness and completely ignores two hidden but critical causal paths. 
\item We introduce a Causal Top-Down Microarchitecture Analysis framework, bridging the semantic gap between algorithmic data treatment and microarchitectural hardware behavior. We open source our workload configuration treatment tools for SPEC CPU2017.
\item We provide empirical evidence of counter-intuitive hardware behaviors across three major microarchitectural platforms: CPI Profiling Blindness, Over-Speculation Congestion, ILP Window Collapse, and Op-Cache SMT Overlap.
\item Based on the ``Ghost State'' and over-speculation discoveries, we offer actionable architectural recommendations, proposing a dynamic speculation throttling mechanism to online throttle pipelines for CPU design.
\end{enumerate}

The remainder of the article is organized as follows. Section~\ref{sec:background} introduces the background and related works. Section~\ref{sec:motivation} discusses the motivation. Section~\ref{sec:methodology} proposes our methodology. Section~\ref{sec:evaluation} shows the evaluation, and Section~\ref{sec:conclusion} concludes and discusses our future work.

\section{Background and Related Work}
\label{sec:background}

To fully comprehend the dynamic interplay between software workloads and hardware pipelines, it is essential to establish the foundational concepts of modern performance analysis and causal inference. This section introduces the terminology and highlights the systemic causal gaps in current benchmarking methodologies.

\subsection{Traditional Benchmarking and Top-Down Analysis}
\textbf{Traditional Benchmarking Paradigms.} Computer architecture heavily relies on empirical methodologies, utilizing benchmark suites~\cite{anderson2011better,panda2018wait,ferdman2012clearing,bienia2008parsec,bienia2008parsec1,woo1995splash,hoste2006comparing,hoste2007analyzing,hoste2006performance,hoste2007microarchitecture,shao2013isa,phansalkar2005measuring,phansalkar2007analysis,limaye2018workload,wang2023wpc,campanoni2010highly} (e.g., SPEC CPU~\cite{SPECCPU89,SPECCPU92,SPECCPU95,SPECCPU2000,SPECCPU2006,SPECCPU2017,SPECCPU2026}, PARSEC~\cite{bienia2008parsec,bienia2011benchmarking,zhan2017parsec3.0}, SPLASH~\cite{woo1995splash}) to proxy real-world performance. The canonical approach aggregates execution outcomes across predefined, static system configurations to yield a synthesized metric (e.g., the SPEC score).

\textbf{The Top-Down Microarchitecture Analysis (TMA).} To understand \textit{why} a CPU achieves a certain score on the given benchmark, the industry standard is Intel's Top-Down Microarchitecture Analysis (TMA)~\cite{yasin2014deep}. By utilizing Hardware Performance Counters (PMUs)~\cite{intelvt,perf,panda2018wait,wang2023wpc,madhav2026spec}, TMA classifies CPU pipeline stalls into a hierarchical tree: Frontend Bound, Backend Bound, Bad Speculation, and Retiring. 
While TMA is a powerful ``speedometer'' that accurately measures the pipeline efficiency (e.g., CPI), it inherently treats the workload configuration as a static entity. 
It provides a snapshot of the hardware's status but lacks the mechanism to trace performance variations back to shifts in the input data or algorithmic adaptations.

\subsection{Bridging Algorithm and Microarchitecture}
To break down the system performance, we must explicitly decouple the software algorithmic behavior from hardware execution efficiency based on the Iron Law of Performance~\cite{hennessy2012computer}: $Elapsed\ Time = \frac{Retired\_Instructions \times CPI}{Frequency}$. We define two key architectural concepts pivotal to our later analysis.

\textbf{Algorithmic Instructions Bloat.} This occurs when a change in the input dataset's distribution (e.g., spatial locality) forces the software algorithm to take a longer execution path (e.g., traversing deeper search trees). The total retired instructions ($Retired\_Ins$) surge exponentially, increasing elapsed time, even if the hardware $CPI$ remains perfectly stable.

\textbf{Capacity-Bound vs. Fill-Rate-Bound Pipelines.} Modern Out-of-Order (OoO) processors rely on deep Reorder Buffers (ROB) to extract Instruction-Level Parallelism (ILP). 
When branch prediction is highly accurate, the CPU can fill a massive ROB (e.g., 256 entries), making performance \textit{Capacity-Bound} (sensitive to the physical size of the buffer). However, under severe control-flow chaos (e.g., extreme data distortion), frequent branch flushes clear the pipeline continuously. The CPU can barely fill a fraction of the ROB before flushing. In this state, the pipeline becomes \textit{Fill-Rate-Bound}; enlarging or halving the physical ROB capacity has minimal effect on overall system performance.

\subsection{Causal Inference and the Object Causal Model}
\label{Sec-Causal}

To rigorously model complex system interactions, we establish our framework on the foundational concepts of cause and effect. Traditional Structural Causal Model (SCM)~\cite{pearl_causality_2000,imbens2015causal,pearl2018book} formalize causal relationships (Cause $A \rightarrow$ Effect $B$) between \textit{variables} $A$ and $B$.  Recent hardware bottleneck quantification frameworks~\cite{dave2023explainable,nasr2025concorde} can be viewed as domain-specific instantiations of these variable-level causal models. However, computer architecture operates across abstracted, hierarchical objects (e.g., Workload, Instructions, and CPU). Purely observation-based, variable-driven SCM often fail to capture interactions across structural depth.

To bridge this semantic gap, Zhan et al.~\cite{zhan2024evaluatology,zhan2025evaluatology} introduced \textbf{Evaluatology} and the \textbf{object causal model (OCM)}. Unlike SCM, OCM centers its modeling on system \textit{objects} (Evaluated Object $EO \Rightarrow$ Affected Object $AO$), characterized by their internal \textit{mechanisms} and \textit{properties (observable metric)}. Within a Self-Contained System (SCS)~\cite{zhan2025evaluatology,wang2026inference}, we can model performance evaluation across four fundamental interacting objects, identifying the dominant causal path among complex interactions:
\begin{enumerate}
    \item \textbf{Workload Object:} Its underlying \textit{algorithm} serves as the mechanism, and the \textit{configuration parameters (e.g., dataset)} represent its observable properties.
    \item \textbf{Instructions Object:} Represents the compiled instruction  execution trace. Its observable property is the \textit{Retired Instructions Count}.
    \item \textbf{CPU Object:} The microarchitecture. Its mechanism is the out-of-order pipeline and microarchitectural behavior (e.g., branch predictor behavior), and its observable properties are the efficiency metric (\textit{CPI}) and behavior metric (e.g., \textit{Branch MPKI}).
    \item \textbf{System Object:} The overarching computer system, with \textit{Performance (Elapsed Time)} as its observable property.
\end{enumerate}

\begin{table}
\centering
\caption{Machine information in our experiments. 
}
\resizebox{\columnwidth}{!}{
\begin{tabular}{llll}
\hline
\textbf{Machine}  & \textbf{Platform A}            & \textbf{Platform B}  & \textbf{Platform C}       \\ \hline
CPU Model Name & Intel Xeon Gold 5120T & AMD EPYC 7543 &  Kunpeng 920 \\
Micro-architecture                &   Skylake       &  Zen3       & Taishan 110                \\
ISA                &   x86-64       &  x86-64       & AArch64                \\
Socket(s)          & 2  & 2                           & 2                          \\
Core(s) Per Socket & 14 & 64                            & 32                         \\
Thread(s) Per Core & 2  & 2                           & 1                          \\
L1 (Per Core)                 & 32KiB I, 32KiB D & 32KiB I, 32KiB D              & 64KiB I, 64KiB D           \\
L2 (Per Core)                & 1MiB                   & 512KiB        & 512KiB                     \\
L3 (Per Socket)                 & 19.25MiB Unified  & 256MiB Unified  & 32MiB Unified  \\
Compiler           & GCC11 & GCC11                    & GCC11                  \\ 
Memory             & 384GiB  & 512GiB             & 384GiB                     \\
Disk               & 2TiB & 252GiB   & 2TiB                       \\
Operating System   & Ubuntu 20.04.6 LTS         & Ubuntu 22.04.5 LTS   & Ubuntu 22.04.4 LTS        \\
Linux kernel   & 5.4.0-216-generic         & 5.15.0-181-generic    & 6.5.0-15-generic          \\
\hline
\end{tabular}
}
\label{Machine-Inf}
\end{table}

\begin{table*}[t]
\centering
\caption{Causal Dataset Treatment Specifications. To guarantee strict ``apples-to-apples'' comparisons, the Absolute Data Scale acts as an invariant anchor across all treatments for each workload. Treatment dimensions (T1, T2) are deliberately dialed to perturb instruction behaviors and CPU microarchitectural states (e.g., Cache, Branch Predictor). Q1--Q4 represent orthogonal extreme quadrants, while Grad-0.1$\sim$0.9 represents smooth parameter interpolations.}
\label{tab:dataset_treatment}
\resizebox{\textwidth}{!}{
\begin{tabular}{@{}l|l|l|l|l|l@{}}
\toprule
\textbf{Workload} &
  \textbf{\begin{tabular}[c]{@{}l@{}}Invariant Data Scale\\ (Anchor to SPEC \texttt{ref})\end{tabular}} &
  \textbf{\begin{tabular}[c]{@{}l@{}}Treatment 1 (T1)\\ $\Rightarrow$ Affected Object\end{tabular}} &
  \textbf{\begin{tabular}[c]{@{}l@{}}Treatment 2 (T2)\\ $\Rightarrow$ Affected Object\end{tabular}} &
  \textbf{\begin{tabular}[c]{@{}l@{}}Orthogonal Extremes\\ (Q1 -- Q4 Matrix)\end{tabular}} &
  \textbf{\begin{tabular}[c]{@{}l@{}}Smooth Gradients\\ (Grad-0.1 $\sim$ 0.9)\end{tabular}} \\ \midrule

\textbf{\begin{tabular}[c]{@{}l@{}}541.leela\_r\\ (MCTS for Go)\end{tabular}} &
  \begin{tabular}[c]{@{}l@{}}Total Moves:\\ 600 Placed Stones\end{tabular} &
  \begin{tabular}[c]{@{}l@{}}\textbf{Spatial Locality} [0.0, 1.0]\\ 0=Global Scatter, 1=Local Fights\\ $\Rightarrow$ \textit{Instructions / CPU}\end{tabular} &
  \begin{tabular}[c]{@{}l@{}}\textbf{Board Size Constraint}\\ 19x19 vs. 9x9 Grid\\ $\Rightarrow$ \textit{Instructions / CPU}\end{tabular} &
  \begin{tabular}[c]{@{}l@{}}\textbf{Q1}: T1=0 (Scattered), T2=19x19\\ \textbf{Q2}: T1=1 (Clustered), T2=19x19\\ \textbf{Q3}: T1=0 (Scattered), T2=9x9\\ \textbf{Q4}: T1=1 (Clustered), T2=9x9\end{tabular} &
  \begin{tabular}[c]{@{}l@{}}Sweep T1 from 0.1 to 0.9\\ while fixing T2=19x19\\ (Observe locality collapse)\end{tabular} \\ \midrule

\textbf{\begin{tabular}[c]{@{}l@{}}505.mcf\_r\\ (Network Simplex)\end{tabular}} &
  \begin{tabular}[c]{@{}l@{}}15,000 Nodes\\ 117,202 Edges\end{tabular} &
  \begin{tabular}[c]{@{}l@{}}\textbf{Spatial Locality} [0.0, 1.0]\\ 0=Scattered, 1=Clustered\\ $\Rightarrow$ \textit{Instructions / CPU}\end{tabular} &
  \begin{tabular}[c]{@{}l@{}}\textbf{Cost Predictability} [0.0, 1.0]\\ 0=Random, 1=Linear\\ $\Rightarrow$ \textit{Instructions / CPU}\end{tabular} &
  \begin{tabular}[c]{@{}l@{}}\textbf{Q1}: T1=0 (Scattered), T2=0 (Random)\\ \textbf{Q2}: T1=0 (Scattered), T2=1 (Linear)\\ \textbf{Q3}: T1=1 (Clustered), T2=0 (Random)\\ \textbf{Q4}: T1=1 (Clustered), T2=1 (Linear)\end{tabular} &
  \begin{tabular}[c]{@{}l@{}}Sweep T2 from 0.1 to 0.9\\ while fixing T1=0\\ (Observe branch sensitivity)\end{tabular} \\ \midrule

\textbf{\begin{tabular}[c]{@{}l@{}}507.cactuBSSN\_r\\ (3D PDE Solver)\end{tabular}} &
  \begin{tabular}[c]{@{}l@{}}Total Volume:\\ $84^3$ Vertices (80 Iters)\end{tabular} &
  \begin{tabular}[c]{@{}l@{}}\textbf{Spatial Distortion} [0.0, 1.0]\\ 0=Perfect Cube, 1=Extreme Needle\\ $\Rightarrow$ \textit{Instructions / CPU}\end{tabular} &
  \begin{tabular}[c]{@{}l@{}}\textbf{N/A (Invariant Physics)}\\ (Topology solely drives bottleneck)\end{tabular} &
  \begin{tabular}[c]{@{}l@{}}\textbf{Q1}: T1=0 (Cube $84{\times}84{\times}84$)\\ \textbf{Q2}: T1=1 (Needle $10{\times}10{\times}5,927$)\\ \textit{Orthogonal comparison achieved via} \\ \textit{extreme geometric warping.}\end{tabular} &
  \begin{tabular}[c]{@{}l@{}}Sweep T1 from 0.1 to 0.9\\ (Observe ILP window collapse)\end{tabular} \\ \midrule

\textbf{\begin{tabular}[c]{@{}l@{}}523.xalancbmk\_r\\ (XML Parsing)\end{tabular}} &
  \begin{tabular}[c]{@{}l@{}}36,000 XML Items\\ ($\sim$58 MB File Size)\end{tabular} &
  \begin{tabular}[c]{@{}l@{}}\textbf{Recursion Depth} [0.0, 1.0]\\ 0=Shallow, 1=Deep (400 chars)\\ $\Rightarrow$ \textit{Instructions / CPU}\end{tabular} &
  \begin{tabular}[c]{@{}l@{}}\textbf{String Randomness} [0.0, 1.0]\\ 0=Repetitive, 1=Random\\ $\Rightarrow$ \textit{Instructions / CPU}\end{tabular} &
  \begin{tabular}[c]{@{}l@{}}\textbf{Q1}: T1=1 (Deep), T2=1 (Random)\\ \textbf{Q2}: T1=1 (Deep), T2=0 (Repetitive)\\ \textbf{Q3}: T1=0 (Shallow), T2=1 (Random)\\ \textbf{Q4}: T1=0 (Shallow), T2=0 (Repetitive)\end{tabular} &
  \begin{tabular}[c]{@{}l@{}}Sweep T1 from 0.1 to 0.9\\ while fixing T2=1\\ (Observe recursion footprint)\end{tabular} \\ \bottomrule
\end{tabular}
}
\end{table*}

\section{Motivation: The Pitfalls of Traditional TMA}
\label{sec:motivation}

\subsection{The Causal Blindness of TMA}
The fundamental goal of CPU design and optimization is to reduce elapsed time. The classic performance equation dictates that $Elapsed\ Time = \frac{Retired\_Instructions \times CPI}{Frequency}$. 

The Top-Down Microarchitecture Analysis (TMA) attempts to optimize this equation by breaking down pipeline stalls to minimize the CPI. However, the community predominantly evaluates TMA using statically prescribed workload configurations (e.g., SPEC CPU \texttt{ref} inputs). This implicitly models performance through a single explicit causal path:
$\textbf{Workload(dataset, algorithm)} \Rightarrow \textbf{CPU(CPI)} \Rightarrow \textbf{System(Performance)}$.
Because TMA only profiles microarchitectural efficiency (the property of the CPU object), it blindly assumes that the workload emits an invariant stream of instructions on the given static configuration. Consequently, performance variations are strictly dictated by how fast the CPU processes that work.

However, from an OCM perspective, this single-path model is fundamentally incomplete and misleading. Real-world applications dynamically adapt to data space properties. Changing the workload  activates two entirely overlooked causal paths that completely bypass or perturb the CPU's CPI.
1) \textbf{The Algorithmic Efficiency Path ($\text{Workload(dataset, algorithm)} \Rightarrow \text{Instructions(Retired)} \Rightarrow \text{System(Performance)}$):} Variations in data topology can dictate the algorithm's execution depth (e.g., algorithmic instruction bloat). This path can double elapsed time while the CPU's CPI remains perfectly stable.
2) \textbf{The Microarchitecture-Algorithm Efficiency  Path ($\text{Workload(dataset, algorithm)} \Rightarrow \text{Instructions(Retired)} \Rightarrow \text{CPU(CPI)} \Rightarrow \text{System(Performance)}$):} An exponential surge in instructions volume can fundamentally alter the execution footprint (e.g., shrinking entirely into the Op-cache). The algorithmic scale dynamically reshapes the hardware state, causing an artificial shift in CPI.

\begin{figure*}[t]
    \centering
    \includegraphics[width=\linewidth]{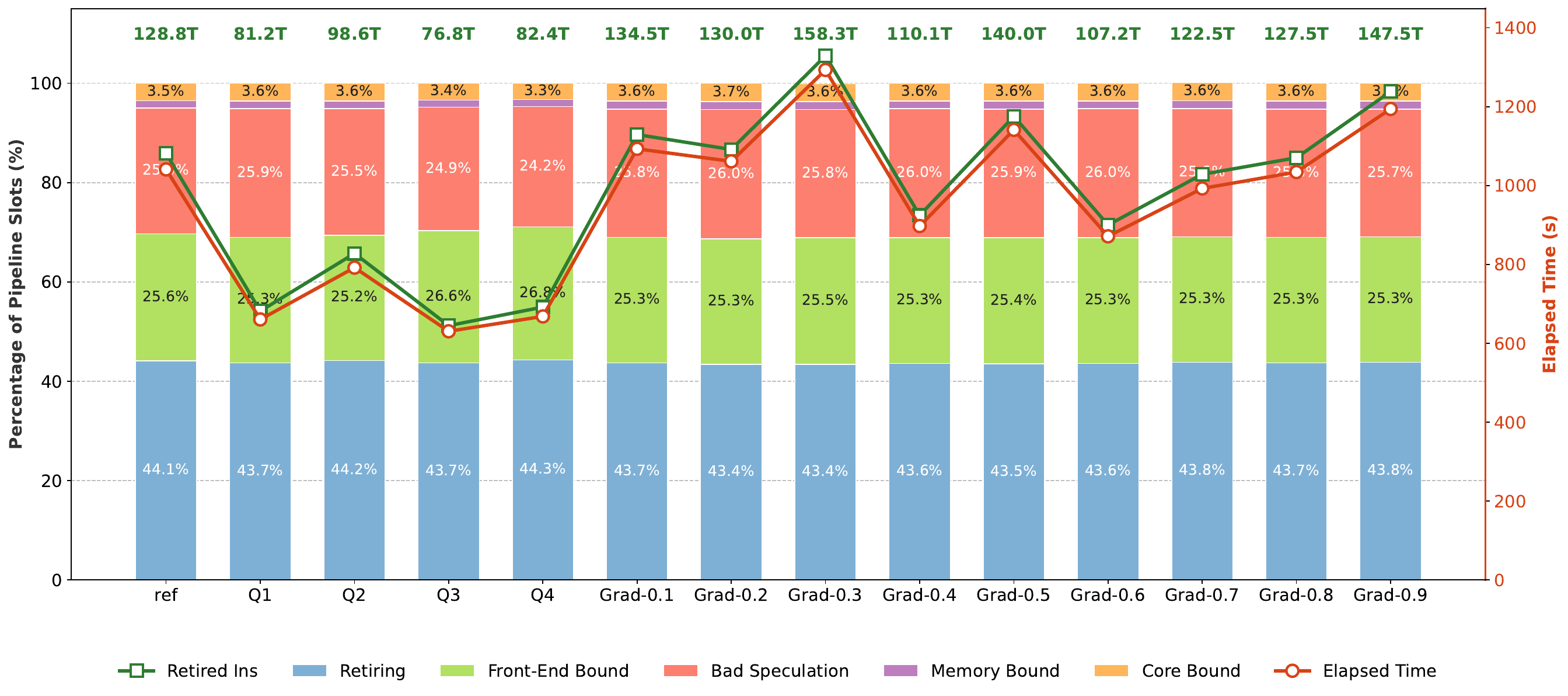}
    \caption{The Microarchitectural Freeze in \textit{leela\_r}: As the dataset's spatial locality varies (Grad-0.1 to 0.9), the TMA pipeline break-down (bar charts) remains entirely frozen. However, the elapsed time (red line) perfectly correlates with the algorithmic instructions bloat (green line), proving the causal path $\text{Workload(dataset, algorithm)} \Rightarrow \text{Instructions(Retired)} \Rightarrow  \text{System(Performance)}$.}
    \label{fig:leela_freeze}
\end{figure*}



In the rest of this section, we apply  treatment to four diverse SPEC CPU2017 workloads on the Intel Gold 5120T Machine in Table~\ref{Machine-Inf}. For each workload, as shown in Table~\ref{tab:dataset_treatment}, we strictly fix the absolute data scale (e.g., total nodes, grid volume, or file size) same as SPEC  \texttt{ref}  to ensure an ``apples-to-apples'' comparison. We then actively mutate specific  treatment (e.g., spatial locality, structural randomness) to generate orthogonal extreme datasets (denoted as Q1-Q4) and smooth parameter gradients (Grad-0.1 to 0.9). By doing so, we activate the TMA-overlooked causal paths ($\textbf{Workload(dataset, algorithm)} \Rightarrow \textbf{Instructions(Retired)} \Rightarrow \textbf{System(Performance)}$ and $\textbf{Workload(dataset, algorithm)} \Rightarrow \textbf{Instructions(Retired)} \Rightarrow \textbf{CPU(CPI)} \Rightarrow \textbf{System(Performance)}$). The following four counter-intuitive case studies demonstrate why traditional TMA leads to causal blindness.

\subsection{The Microarchitectural Freeze (\textit{leela\_r})}\label{sec:mot:541}
\textbf{Dataset Treatment Strategy:} In Monte Carlo Tree Search (\textit{leela\_r}), we generated treatment SGF datasets by strictly fixing the total moves on Go board (600 moves). We exclusively manipulated the \textit{spatial locality} gradient. Stones were progressively shifted from being globally scattered across the board to being locally clustered, spanning the orthogonal extreme quadrants (Q1--Q4) and smooth parameter interpolations (Grad-0.1$\sim$0.9) defined in Table~\ref{tab:dataset_treatment}.

\textbf{The Illusion of Profiling:} The most dangerous flaw of TMA is its causal blindness to algorithmic instructions bloat. As shown in Figure~\ref{fig:leela_freeze}, the TMA pipeline breakdown (background bars) remains entirely ``frozen'' across all  treatments. The Retiring ($\sim$44\%), Frontend Bound ($\sim$25\%), and Bad Speculation ($\sim$26\%) slots are virtually identical, and the overall CPI stabilizes at approximately 1.17.

\textbf{The Causal Reality} ($\textbf{Workload(dataset, algorithm)} \Rightarrow \textbf{Instructions(Retired)} \Rightarrow \textbf{System(Performance)}$): A computer architect looking solely at these TMA snapshots would confidently conclude that the workload's behavior is invariant. Yet, the elapsed time  (red line) oscillates violently from 630s to 1293s ($>100\%$ variation). Clustering the data locality did not alter the pipeline efficiency (CPI); instead, it trapped the MCTS algorithm in deep tactical evaluations, inflating the number of retired instructions (green line) from 76.8T to 158.3T. TMA only shows how efficiently the processor is running, completely ignoring that the algorithm forced the CPU to execute more than twice the amount of work.

\begin{figure}[t]
    \centering
    \includegraphics[width=\linewidth]{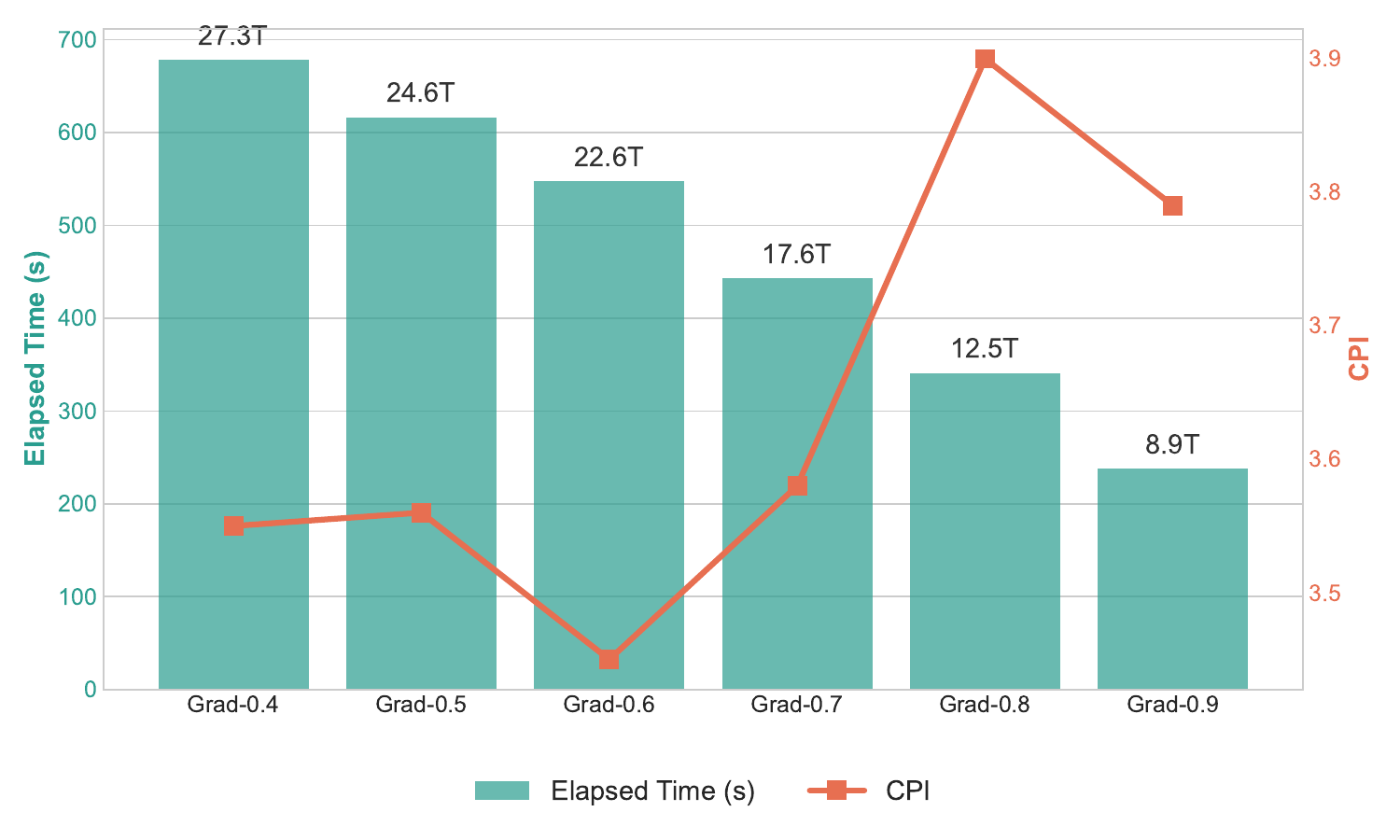}
    \caption{The Pointer-Chasing Paradox in \textit{mcf\_r}: As graph edges become more structured, CPI drastically worsens, yet overall elapsed time counter-intuitively drops.}
    \label{fig:mcf_paradox}
\end{figure}

\subsection{The Pointer-Chasing Paradox (\textit{mcf\_r})}
\textbf{Dataset Treatment Strategy:} In the network simplex algorithm (\textit{mcf\_r}), we maintained a strict graph scale (15,000 nodes and 117,202 edges) but modulated the \textit{cost randomness}. As shown in Table~\ref{tab:dataset_treatment}, we smoothly shifted the edge costs from purely random values to a predictable, linear distance-based pattern.

\textbf{The Directional Fallacy:} Even when TMA accurately detects a change in CPI, blindly optimizing for it can lead to paradoxical results. Figure~\ref{fig:mcf_paradox} illustrates a striking anomaly among the Grad-0.6 and Grad-0.8 datasets: the CPI worsens drastically (from 3.4 to 3.9), but the elapsed time drops precipitously from 550s to 350s.

\textbf{The Causal Reality ($\text{Workload(dataset, algorithm)} \Rightarrow \text{Instructions(Retired)} \Rightarrow  \text{System(Performance)}$ overriding $\text{Workload(dataset, algorithm)} \Rightarrow \text{CPU(CPI)} \Rightarrow \text{System(Performance)}$):}
TMA would flag Grad-0.8 as a microarchitectural disaster. The predictable edge costs ironically turn highly accurate branch predictions into ``Over-Speculation Congestion,'' where the CPU aggressively fetches deep, dependent pointers, causing severe Line Fill Buffer (LFB) congestion. However, the structured data also allows the algorithm to find shorter routing paths, plummeting the instructions count from 22.6T to 12.5T. The massive reduction in algorithmic work entirely overrides the severe microarchitectural penalty. Optimizing the CPU pipeline solely based on the bloated CPI of Grad-0.8 would be fundamentally misguided.

\begin{figure}[t]
    \centering
    \includegraphics[width=\linewidth]{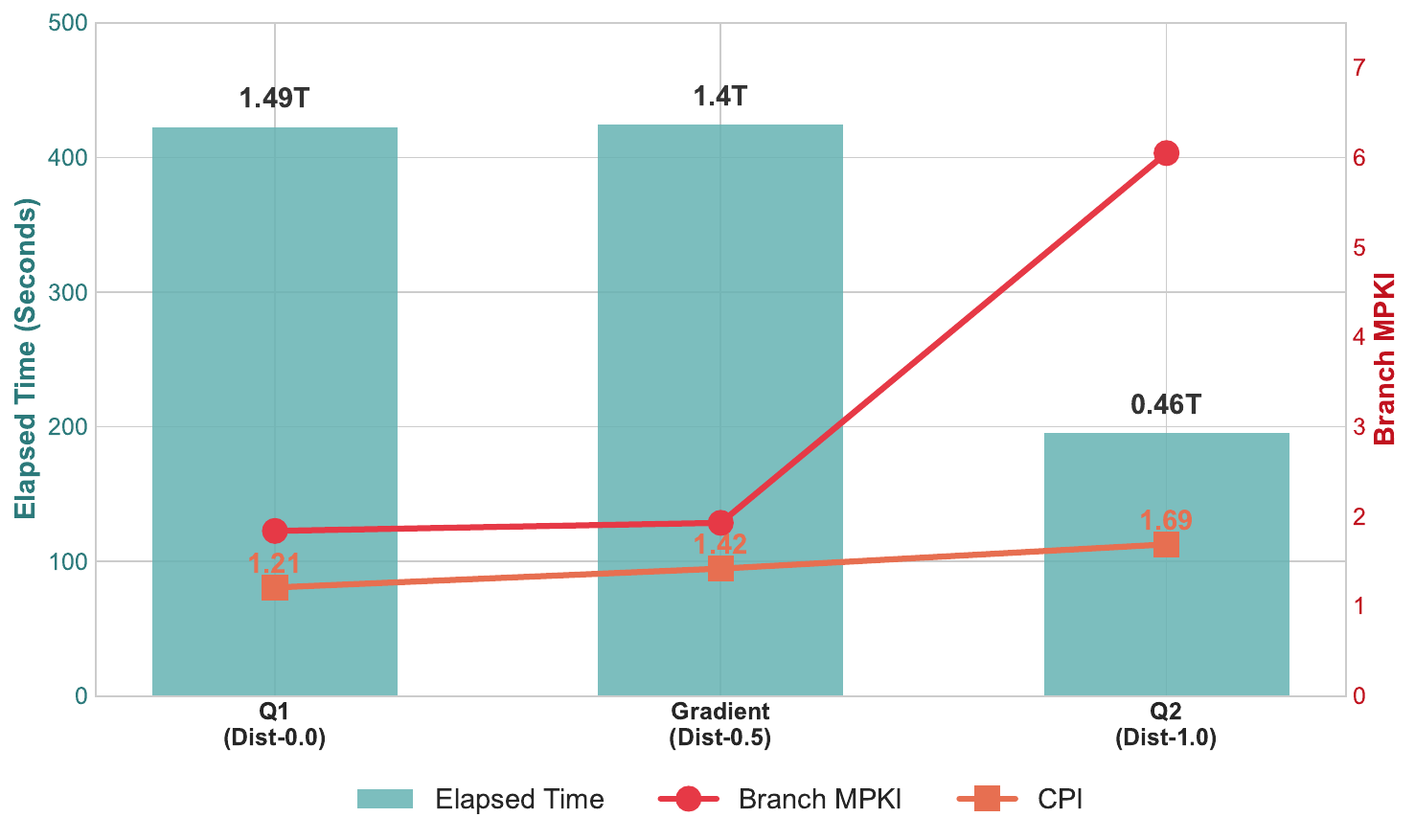}
    \caption{The Masked Disaster in \textit{cactuBSSN\_r}: Algorithmic pruning causes elapsed time to drop, completely hiding the exponential surge in Branch MPKI.}
    \label{fig:cactuBSSN_disaster}
\end{figure}

\subsection{The Hidden Bottleneck Shift (\textit{cactuBSSN\_r})}
\textbf{Dataset Treatment Strategy:} 
Following Table~\ref{tab:dataset_treatment},
in the 3D PDE solver (\textit{cactuBSSN\_r}), we kept the total computational volume strictly identical ($84^3$ vertices) but morphed the \textit{spatial distortion}. We stretched the space from a perfect cube (Q1, $84 \times 84 \times 84$) to an intermediate form (Gradient-0.5, $47 \times 47 \times 268$), and  then to an extreme needle-shape (Q2, $10 \times 10 \times 5,927$).

\textbf{The Masked Disaster:} Extreme spatial distortion results in a massive 70\% drop in retired instructions (from 1.47T down to 0.45T) due to algorithmic boundary pruning and early PDE convergence. Naturally, the elapsed time drops, giving the illusion of a lighter and healthier workload (Figure~\ref{fig:cactuBSSN_disaster}). However, this reduction in time hides a catastrophic shift in the underlying bottleneck. Under extreme distortion, short innermost loops ($10 \times 10$) destroy hardware prefetcher strides and cause the Branch MPKI to triple (from 1.84 to 6.05). The pipeline abruptly shifts from being \textit{capacity-bound} (efficiently filling a 256-entry Reorder Buffer) to \textit{fill-rate-bound} (stalling constantly due to branch flushes). 


\textbf{The Causal Reality (The Tug-of-War between $\text{Workload(dataset, algorithm)} \Rightarrow \text{Instructions(Retired)} \Rightarrow  \text{System(Performance)}$ and $\text{Workload(dataset, algorithm)} \Rightarrow \text{CPU(CPI)} \Rightarrow  \text{System(Performance)}$):} 
This workload exposes a dynamic ``Tug-of-War'' between two causal paths. When shifting from Q1 to the extreme Q2, the algorithmic efficiency path ($\text{Workload(dataset, algorithm)} \Rightarrow \text{Instructions(retired)} \Rightarrow \text{System(Performance)}$) is the absolute dictator; the 70\% instructions reduction easily masks the microarchitectural penalty, dropping elapsed time from 422.6s to 195.2s. Conversely, the transition from Q1 to Gradient-0.5 reveals the power of the microarchitectural efficiency path ($\text{Workload(dataset, algorithm)} \Rightarrow \text{CPU(CPI)} \Rightarrow \text{System(Performance)}$). As the space begins to stretch, instructions drop slightly (from 1.49T to 1.40T), yet the elapsed time paradoxically \textit{increases} from 422.6s to 424.4s. The geometric distortion hits a pathological threshold for the pipeline Frontend, driving the CPI up from 1.21 to 1.42, overpowering the algorithmic reduction. TMA is fundamentally unequipped to capture this non-linear, threshold-driven reality.

\begin{figure}[t]
    \centering
    \includegraphics[width=\linewidth]{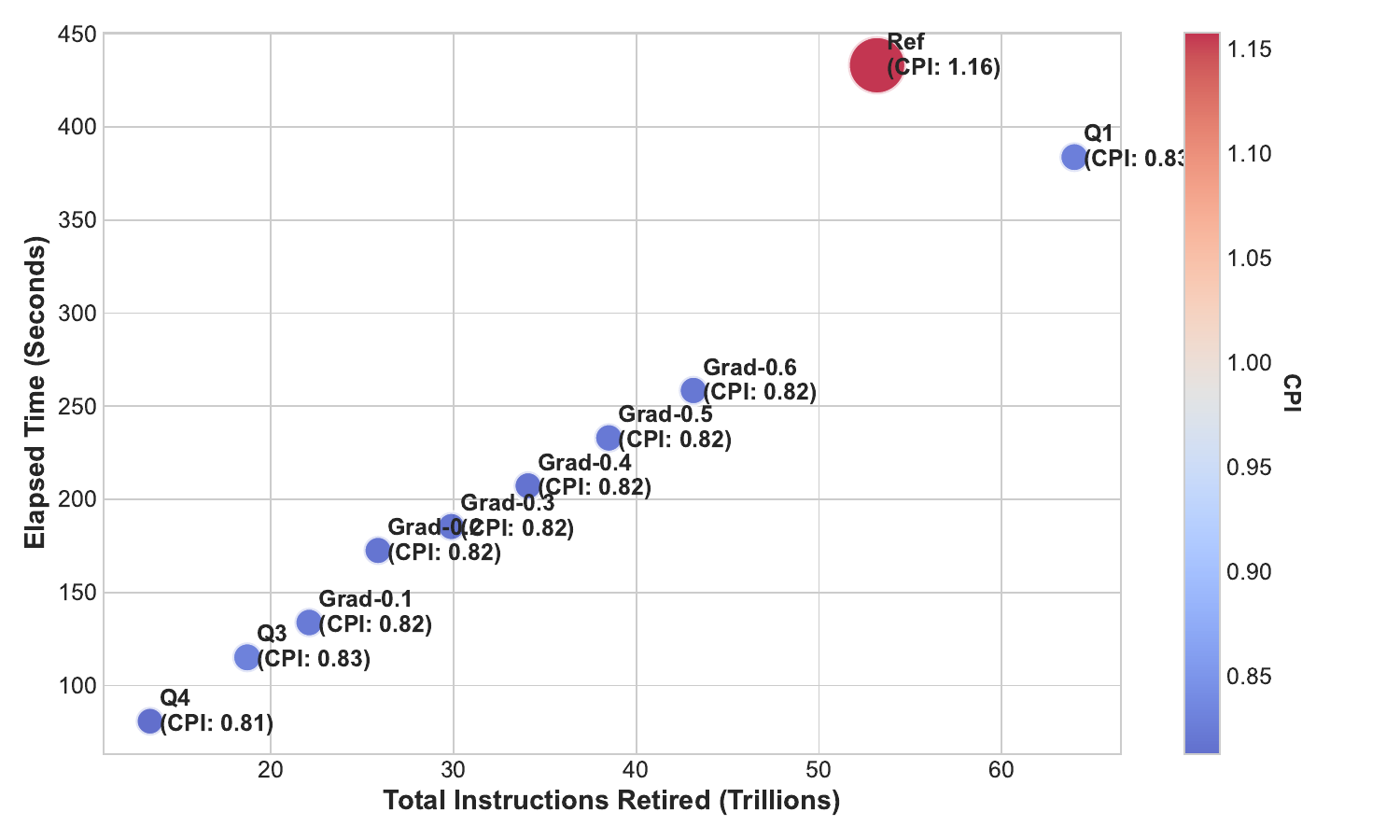}
    \caption{The Workload Anomaly in \textit{xalancbmk\_r}: The  retiring 65 Trillion instructions finishes faster than the reference  retiring only 53 Trillion instructions.}
    \label{fig:xalancbmk_anomaly}
\end{figure}

\subsection{The Workload Anomaly (\textit{xalancbmk\_r})}\label{sec:mot:523}
\textbf{ Treatment Strategy:} As detailed in Table~\ref{tab:dataset_treatment}, in XML parsing (\textit{xalancbmk\_r}), we fixed the input file item at 36,000 items. To trigger varying degrees of nested XSLT template calls, we progressively increased the \textit{recursion depth} by padding inner string lengths from 2 to 400 characters.

\textbf{The Bandwidth Illusion:}  
Through causal treatment, we observed a complex bottleneck drift to a significant Frontend Bound (the break-down Frontend bound shifts from 20.2\% on SPEC \texttt{ref} dataset to more than 32\% on our treatment datasets). Moreover, as shown in Figure~\ref{fig:xalancbmk_anomaly}, the Q1 dataset retires 65 Trillion instructions, yet finishes faster than the SPEC \texttt{ref} dataset, which retires only 53 Trillion instructions.

\textbf{The Causal Reality ($\text{Workload(dataset, algorithm)} \Rightarrow \text{Instructions(Retired)}  \Rightarrow \text{CPU(CPI)} \Rightarrow \text{System(Performance)}$):} 
This anomaly exposes a compounded causal interaction. Deep recursion triggers a massive instructions bloat via nested \texttt{CALL/RET} operations. However, this specific bloat collapses the dynamic instructions footprint into an ultra-small code segment. This perfectly packs the execution into the Op-cache (DSB), completely bypassing the legacy decode pipeline and saturating the Frontend dispatch bandwidth. Consequently, the CPI drops so dramatically (from 1.16 down to 0.83) that it overcompensates for the extra algorithmic work.


\subsection{Summary: The Need for Causal TMA}


These four cases demonstrate that TMA snapshots suffer from severe causal blindness. This blindness is why relying solely on TMA is dangerous---it inevitably misleads architects into optimizing hardware for what are essentially software anomalies (e.g., blindly expanding ROB capacity for \textit{cactuBSSN\_r} when it is fill-rate-bound). 

To capture the holistic performance landscape, we propose the \textbf{Causal Top-Down Microarchitecture Analysis (CTMA)}. By deliberately intervening on the Workload object and deploying hardware probes to cross-validate on the CPU object, CTMA explicitly models and validates all three causal paths, providing a robust, generalized framework for CPU design.

\section{The Causal Top-Down Microarchitecture Analysis (CTMA)}
\label{sec:methodology}

As demonstrated in Section~\ref{sec:motivation}, traditional TMA suffers from causal blindness when algorithmic-related causal paths dynamically adapt. To systematically decouple these complex system interactions, we propose \textbf{Causal Top-Down Microarchitecture Analysis (CTMA)}. 



\begin{figure*}[t]
    \centering
    \includegraphics[scale=0.9]{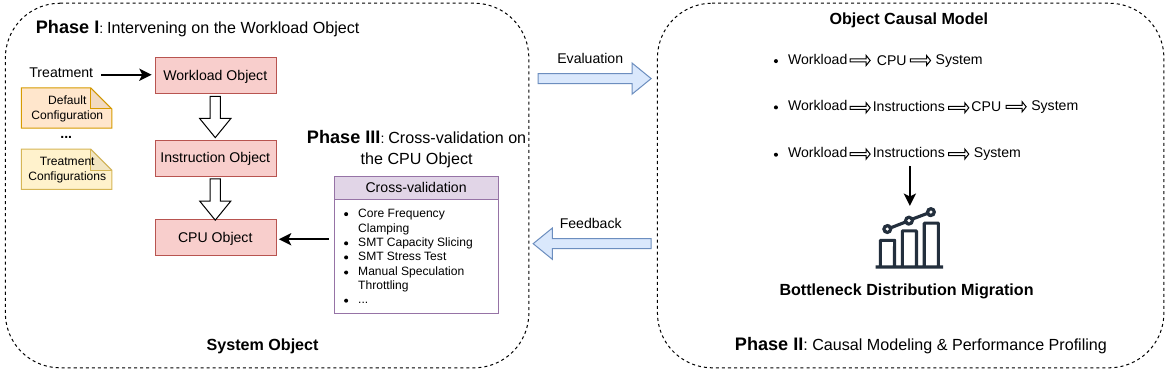}
    \caption{Overview of the Causal Top-Down Microarchitecture Analysis (CTMA). By treating the workload configuration and perturbing microarchitectural states via hardware probes, we capture dynamic bottleneck migrations and cross-validate causal performance models. The framework accommodates both real-silicon probes and cycle-accurate simulators.}
    \label{fig:framework}
\end{figure*}

Rather than a rigid set of specific experimental steps, CTMA is designed as a generalized, plug-and-play framework based on the object causal model (OCM). As illustrated in Figure~\ref{fig:framework}, it establishes a tight feedback loop between System object and a casual modeling and performance profiling engine. Instead of passively profiling a static workload, CTMA explicitly defines two abstract treatment interfaces: the \textbf{Software Interface} (intervening on the \textit{Workload Object}) and the \textbf{Hardware Interface} (intervening on the \textit{CPU Object}). This framework identifies and models dynamic bottleneck migrations and rigorously cross-validates the modeled causal paths. CTMA operates in three fundamental phases.

\subsection{Phase I: The Software Interface (Intervening on the Workload Object)}
\label{sec:phase1_software}
The first phase aims to activate the hidden causal paths overlooked by TMA by systematically intervening the \textbf{Workload Object}. In the OCM paradigm, a workload's underlying algorithm acts as its mechanism, and its configuration parameters act as its properties. Therefore, the Software Interface is designed to mutate the workload properties (e.g., s, compiler flags, concurrency) to trigger dynamic algorithmic adaptations.

To ensure scientific rigor, this phase operates strictly around an anchor point: the \textbf{``Default Configuration''} (e.g., the official SPEC \texttt{ref} inputs). We generate a spectrum of \textbf{``Treatment Configurations''} by dialing specific workload properties (e.g., spatial locality of dataset) known to change algorithmic instructions behaviors and stress microarchitectural bottlenecks, while strictly preserving other properties same with the default.

\textbf{Our Reference Implementation:} In this article, we implement the Software Interface via \textit{DataSet Treatment}. To maintain the ``apples-to-apples'' anchor, we fix the absolute data scale (e.g., maintaining exactly $84^3$ vertices in \textit{cactuBSSN\_r}, or 600 moves in \textit{leela\_r}). By keeping the total scale constant while morphing its topological distribution, any fluctuation in the retired instructions is definitively isolated as an algorithmic adaptation (e.g., boundary pruning), rather than a trivial change in raw input size.




\subsection{Phase II: Dynamic Bottleneck Analysis (Using OCM for Causal Modeling and Performance Profiling)}
\label{sec:phase2_modeling}
Once the treated Workload Objects are compiled to Instructions Objects and applied to the CPU Objects, the resulting computer system (System Objects) yield performance profiles. Then CTMA  transitions to causal modeling (right side of Figure~\ref{fig:framework}). Unlike traditional TMA that collapses all system behaviors into an opaque overall system metric (CPI or Elapsed Time), CTMA utilizes the object causal model (OCM) to track interactions across the four fundamental system objects: Workload, Instructions, CPU, and System.

\begin{figure*}[t]
    \centering
    \includegraphics[scale=0.439]{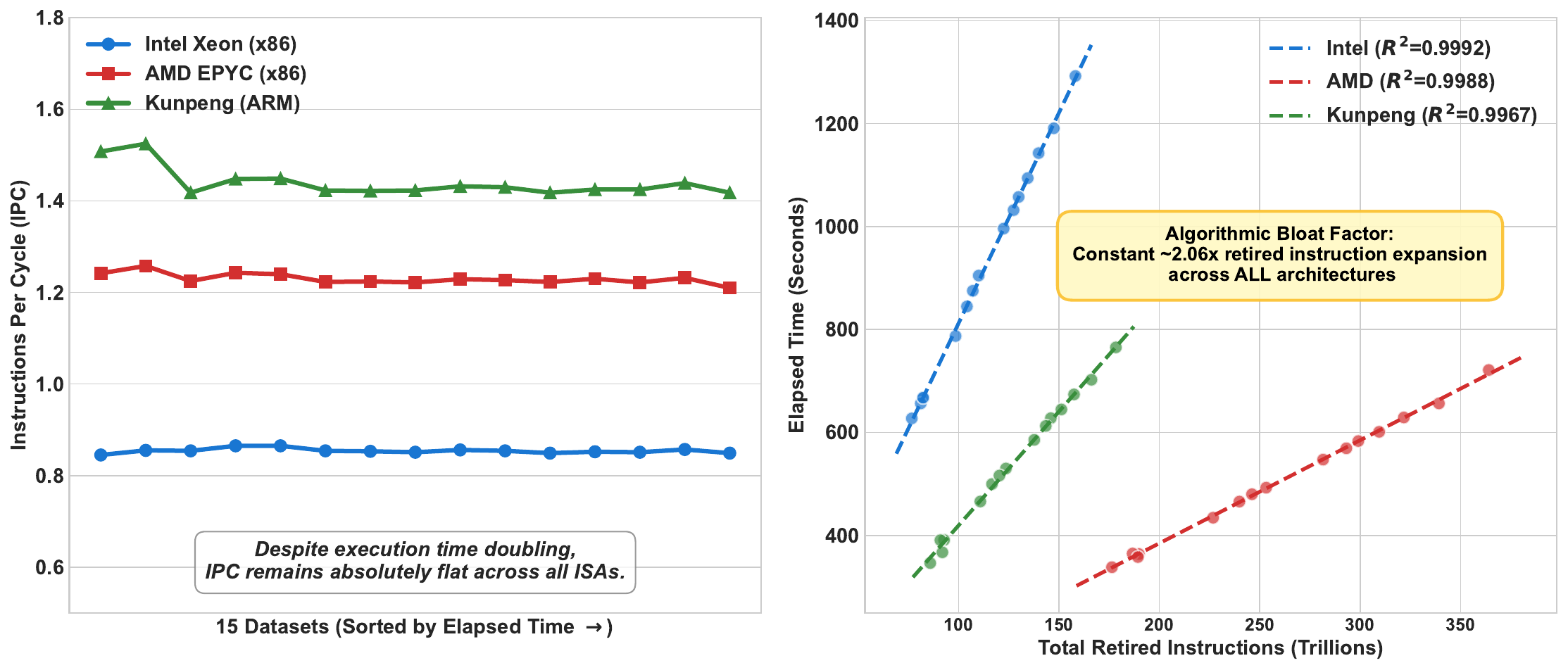}
    \caption{The ``Microarchitectural Freeze'' in \textit{leela\_r} across three major commercial platforms. Elapsed time doubles strictly due to the Instructions Object (Algorithmic Bloat), while the CPU Object (TMA pipeline slots and CPI) remains utterly blind to the degradation.}
    \label{fig:leela_three}
\end{figure*}

By mapping the bottleneck migration across treated configurations, CTMA  models three distinct causal paths.
1)  \textbf{$\text{Workload(dataset, algorithm)} \Rightarrow \text{CPU(CPI)} \Rightarrow  \text{System(Performance)}$ }: The traditional microarchitectural efficiency path. Variations in input data directly alter hardware behavior (e.g., spatial distortion destroying prefetcher strides).
2) \textbf{$\text{Workload(dataset, algorithm)} \Rightarrow \text{Instructions(Retired)} \Rightarrow  \text{System(Performance)}$}: The algorithmic efficiency path. This formalizes how algorithmic instructions bloat dictates elapsed time, successfully explaining why performance plummets even when pipeline efficiency (CPI) remains completely frozen (as seen in Figure~\ref{fig:leela_freeze} and Figure~\ref{fig:leela_three}).
3) \textbf{$\text{Workload(dataset, algorithm)} \Rightarrow \text{Instructions(Retired)}  \Rightarrow \text{CPU(CPI)} \Rightarrow \text{System(Performance)}$}: The microarchitecture-algorithm efficiency path. Changes in instructions volume can fundamentally alter the execution footprint (e.g., shrinking entirely into the Op-cache), thereby artificially inflating the CPU's pipeline efficiency.
These decoupled causal paths provide a robust, mathematical blueprint of the real-world system constraints, transcending the ``black-box'' nature of traditional TMA under the static workload configuration. 

\subsection{Phase III: The Hardware Interface (Cross-Validation on the CPU Object)}
\label{sec:phase3_hardware}
The final phase addresses a profound challenge in systems research: validating theoretical model predictions against physical execution. To prevent correlational illusions, CTMA implements a ``Cross-validation'' loop using \textbf{Treatment Hardware Probes}. 

As a generalized plug-and-play framework, the Hardware Interface of CTMA accepts any mechanism capable of probing or constraining the \textit{CPU Object}. Researchers can plug in cycle-accurate simulators~\cite{binkert2006m5,martin2005multifacet,bartholomew2006qemu,li2009mcpat,patel2011marss,binkert2011gem5} (e.g., gem5~\cite{binkert2011gem5}) for ultimate register-level control, or utilize real silicon environments. 

\textbf{Our Reference Implementation:} To prove the ecological validity of our findings, we forgo simulators and deploy a suite of \textbf{Non-Invasive Real-Silicon Probes} across commercial processors (Intel, AMD, ARM in Table~\ref{Machine-Inf}). Instead of introducing confounding variables via artificial noise code (e.g., cache-thrashing memory loops), 
we exploit native hardware facilities or controlled software codes to enforce constraints on the CPU Object:

\begin{itemize}
    \item \textbf{Core Frequency Clamping:} To eliminate the non-deterministic noise of modern power management (e.g., thermal throttling and Turbo Boost) and to isolate memory-wall effects, we deploy OS-level frequency governors (e.g., \texttt{cpupower}). By explicitly disabling Turbo Boost and clamping the CPU to a strict frequency (e.g., 1.2 GHz), we forcefully decouple the processor's compute speed from the uncore memory frequency. 
    \item \textbf{SMT Resource Partitioning:} To differentiate between \textit{capacity-bound} and \textit{fill-rate-bound} states, we pin a highly constrained process (e.g., executing only \texttt{\_mm\_pause()}) to the sibling logical core. This triggers the hardware to statically or dynamically partition the Reorder Buffer (ROB) and queues without consuming dynamic fetch bandwidth, cleanly isolating the impact of Backend depth.
    \item \textbf{SMT Stress Test:} To expose Frontend resource sharing policies, we co-run identical processes of a target workload aggressively pinned to a pair of sibling logical cores (e.g., utilizing OS-level affinity via \texttt{taskset -c 1} and \texttt{29} on the Intel Xeon Machine). Compared with a solitary execution with the sibling logical core idle, this concurrent setup forces the CPU to dynamically arbitrate fetch bandwidth. It cleanly reveals whether the hardware achieves cycle-level overlapping (e.g., seamlessly interleaving fetch slots from a shared Decoded Stream Buffer during \texttt{CALL} stalls) or suffers massive performance regressions due to rigid Op-cache capacity partitioning.
    \item \textbf{Manual Speculation Throttling:} To verify ``Over-Speculation Congestion,'' we dynamically patch hot functions with explicit \texttt{NOP} instructions on the workload source code. This acts as a hardware ``traffic light,'' artificially slowing the Frontend to observe if the Backend memory queue congestion alleviates and performance paradoxically improves.
\end{itemize}

By aligning the observed shifts triggered via the Software Interface (Workload Object) with the deterministic responses from the Hardware Interface (CPU Object), CTMA achieves fully closed-loop causal verification. This generalizable framework provides architects with actionable, dynamic insights, fundamentally redefining the performance profiling and CPU design space exploration (DSE) paradigm.

\section{Evaluation}
\label{sec:evaluation}

In this section, we deploy the hardware probes defined in our methodology (Section~\ref{sec:phase3_hardware}) to cross-validate the causal bottleneck models (Section~\ref{sec:phase2_modeling}). We execute our CTMA across three distinct microarchitectures to prove the universality of our findings.

\subsection{Experimental Setup}
As shown in Table~\ref{Machine-Inf}, our evaluation spans three major commercial platforms: Intel Xeon Gold 5120T (Skylake-SP), AMD EPYC 7543 (Zen 3), and ARM Kunpeng 920 (ARMv8).

To rigorously eliminate Operating System (OS) scheduling noise and background interrupts, we explicitly isolate the testing cores using GRUB kernel parameters. For instance, on the Intel Xeon machine, we apply \texttt{GRUB\_CMDLINE\_LINUX="isolcpus=1,29 nohz\_full=1,29 rcu\_nocbs=1,29 irqaffinity=0,28"}. The baseline SPEC process is strictly pinned to an isolated logical core (e.g., core 1). When an SMT hardware probe is applied, the treatment process (e.g., the \texttt{\_mm\_pause()} workload) is pinned to the isolated sibling logical core (e.g., core 29 on the Intel Xeon Machine). Hardware performance events are collected via Linux \texttt{perf}~\cite{perf} and Intel VTune~\cite{intelvt}.


\subsection{Validating Algorithmic Instruction Bloat via Frequency Clamping (\textit{leela\_r})}
\textbf{Software Treatment Insight:} As established in Section~\ref{sec:motivation}, manipulating spatial locality in \textit{leela\_r} triggers a ``Microarchitectural Freeze.'' The TMA CPI profile remains entirely static, yet elapsed time oscillates violently, driven by a $2.06\times$ constant instructions  bloat across all three platforms (Figure~\ref{fig:leela_three}). To definitively eliminate the hypothesis of a hidden ``Memory Wall'' (which TMA might misclassify), we deploy the \textbf{Core Frequency Clamping} hardware probe.

\textbf{Hardware Probe Cross-Validation:} If a workload is memory-bound, downclocking the CPU core will \textit{not} proportionally increase elapsed time, as the uncore RAM frequency remains constant. Table~\ref{tab:freq-scaling} demonstrates the elapsed time under high and low frequency locks. Across the Intel, ARM, and AMD platforms, the elapsed time ratio perfectly matches the physical frequency ratio (e.g., $2.2GHz / 1.2GHz \approx 1.833$) across all  treatments. This linear scaling provides strong evidence that the CPU remains in a steady compute-bound state, fully executing the bloated instructions. 

\begin{table*}[t]
\centering
\caption{Frequency Clamping Validates Pure Instruction Bloat (\textit{leela\_r})}
\label{tab:freq-scaling}
\resizebox{\linewidth}{!}{
\begin{tabular}{l|cc|c|cc|c|cc|c}
\toprule
\multirow{2}{*}{\textbf{Dataset}} & \multicolumn{3}{c|}{\textbf{Intel Xeon (2.2GHz vs 1.2GHz)}} & \multicolumn{3}{c|}{\textbf{ARM Kunpeng (2.2GHz vs 1.2GHz)}} & \multicolumn{3}{c}{\textbf{AMD EPYC (2.8GHz vs 2.1GHz)}} \\ \cline{2-10} 
 & Time (2.2) & Time (1.2) & \textbf{Scaling Ratio}
 & Time (2.2) & Time (1.2) & \textbf{Scaling Ratio}
 & Time (2.8) & Time (2.1) & \textbf{Scaling Ratio} \\ \midrule
\texttt{ref} & 1222.2s & 2243.6s & \textbf{1.836x} & 706.0s & 1289.9s & \textbf{1.827x} & 669.2s & 894.6s & \textbf{1.337x} \\
\textbf{Q1} & 784.8s & 1421.2s & \textbf{1.811x} & 458.8s & 841.9s & \textbf{1.835x} & 430.5s & 579.9s & \textbf{1.347x} \\
\textbf{Q2} & 939.1s & 1725.3s & \textbf{1.837x} & 548.0s & 1001.8s & \textbf{1.828x} & 514.1s & 685.0s & \textbf{1.332x} \\
\textbf{Q3} & 740.2s & 1375.1s & \textbf{1.858x} & 407.1s & 745.3s & \textbf{1.831x} & 398.7s & 532.4s & \textbf{1.335x} \\
\textbf{Q4} & 784.4s & 1443.4s & \textbf{1.840x} & 433.6s & 788.9s & \textbf{1.819x} & 423.1s & 562.3s & \textbf{1.329x} \\ \midrule
\textit{Expected} & \multicolumn{2}{c|}{$2.2 / 1.2$} & \textbf{1.833x} & \multicolumn{2}{c|}{$2.2 / 1.2$} & \textbf{1.833x} &  \multicolumn{2}{c|}{$2.8 / 2.1$} & \textbf{1.333x} \\ \bottomrule
\end{tabular}
}
\end{table*}

\subsection{Isolating Backend Vulnerabilities via SMT Slicing (\textit{cactuBSSN\_r})}
\textbf{Software Treatment Insight:} In \textit{cactuBSSN\_r}, changing spatial distortion from a perfect cube (Q1) to an extreme needle-shape (Q2) causes the elapsed time to drop significantly. However, this algorithmic pruning masks a severe microarchitectural penalty: Branch MPKI surges drastically (e.g., from 1.84 to 6.05 on Intel, and 0.11 to 0.19 on AMD), causing Backend stalls to explode. We causally model this as a transition from a capacity-bound state to a fill-rate-bound state. 

\textbf{Hardware Probe Cross-Validation:} To prove this, we deploy the \textbf{SMT Resource Partitioning} probe on the x86 platforms (as Kunpeng 920 does not support SMT). By co-running a \texttt{\_mm\_pause()} process on the sibling logical core, we partition the physical ROB entries. On Intel, the ideal topology (Q1) degrades by $\sim$23\% under ROB partitioning, confirming its reliance on a massive instructions window to extract ILP (Capacity-Bound). Conversely, the extremely distorted topology (Q2) degrades by only $\sim$9\% (AMD exhibits a similar $\sim$12\% vs. $\sim$10\% trend). Because Q2 suffers from severe branch flushes, it can never fill even a partitioned ROB. 
This hardware probe provides targeted evidence of the fill-rate vulnerability that TMA overlooks.

\subsection{The Architectural Divide of Op-Cache Sharing (\textit{xalancbmk\_r})}
\textbf{Software Treatment Insight:} Deep recursion in \textit{xalancbmk\_r} causes a massive instructions surge ($13.4T \to 308.2T$ on Intel, $15.7T \to 382.0T$ on Kunpeng, and $39.2T \to 677.1T$ on AMD). Counter-intuitively, ICache misses remain negligible across all platforms (Max L1 ICache Miss: 2.84\% on Intel, 0.99\% on Kunpeng, 1.25\% on AMD), yet the Frontend Bandwidth Bound skyrockets to 24\% on Intel. The dense \texttt{CALL/RET} operations shatter the fetch bandwidth, trapping the execution entirely within a tiny code footprint inside the Op-cache.

\begin{figure}[t]
    \centering
    \includegraphics[width=\linewidth]{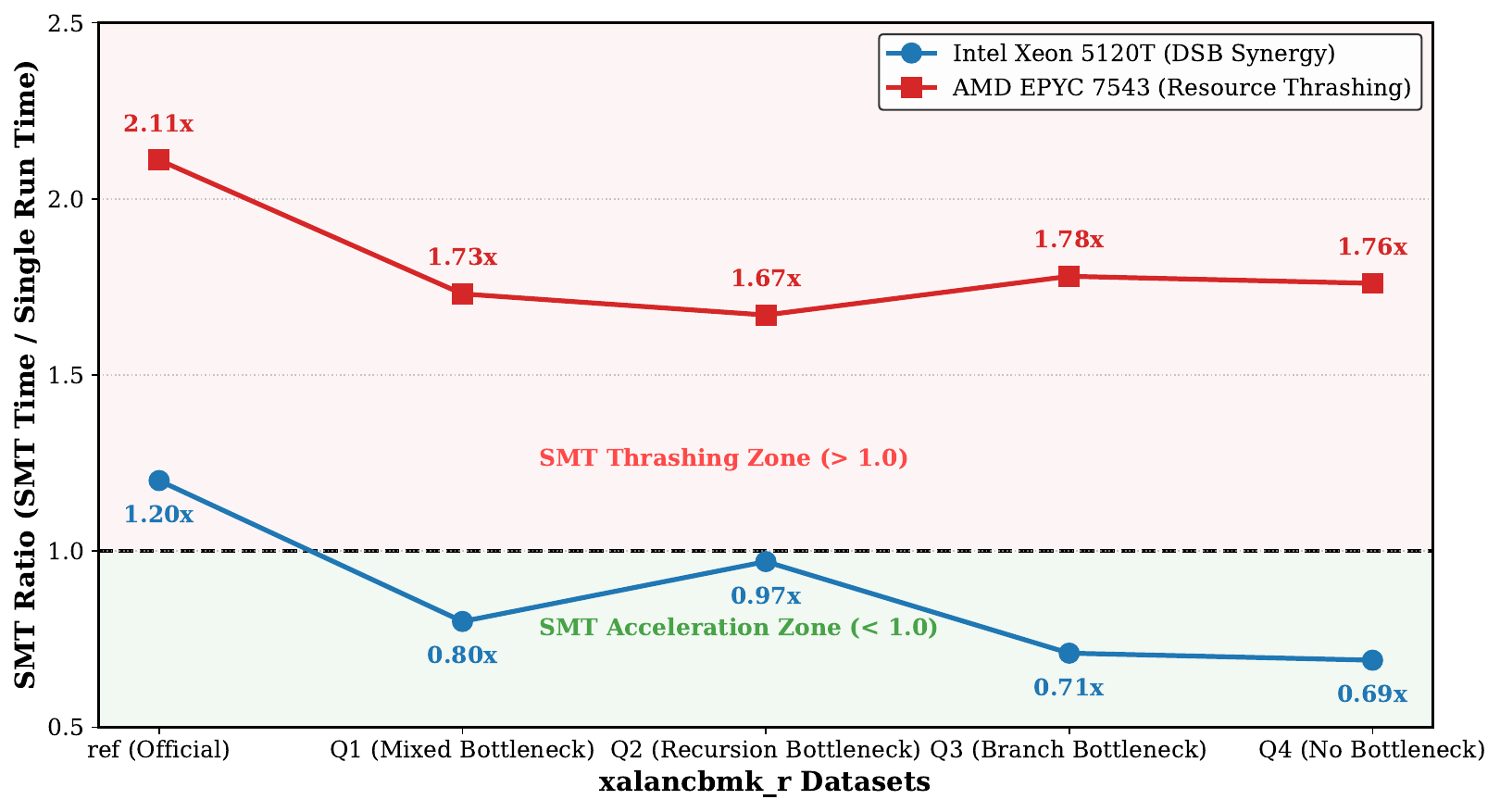}
    \caption{SMT Stress Test (\textit{xalancbmk\_r}): Intel achieves speedups due to DSB overlap on small footprints, while AMD suffers severe resource thrashing.}
    \label{fig:xalancbmk_smt}
\end{figure}

\textbf{Hardware Probe Cross-Validation:} To verify this ``Bandwidth Illusion,'' we deploy the \textbf{SMT Stress Test}. As shown in Figure~\ref{fig:xalancbmk_smt}, Intel surprisingly achieves  speedups. Because the tiny code footprint allows a large fraction of the $\mu$ops to be delivered from Intel's Decoded Stream Buffer (DSB), for example, 53\% of the $\mu$ops are delivered from the DSB on the Q2.
The two processes exhibit constructive cycle-level overlapping: when Process A encounters control-flow or Frontend stalls, Process B can utilize otherwise idle fetch and execution slots. This SMT-enabled latency hiding allows each workload instance to complete faster under co-execution than when running alone. 
Conversely, AMD Zen 3 suffers a massive performance regression, indicating that its rigid Op-cache partitioning policies fail to exploit this ultra-small footprint overlap. 

\subsection{Throttling Over-Speculation Congestion via NOP Injection (\textit{mcf\_r})}

\textbf{Software Treatment Insight:} To rigorously infer the effects of branch predictability and spatial locality on the microarchitecture, we generated four treatment datasets (Q1--Q4) for \textit{mcf\_r} maintaining the exact same graph scale (nodes and edges) as the official \texttt{ref} input. We paired them by spatial locality: Q1 and Q2 share scattered spatial locality, whereas Q3 and Q4 share clustered spatial locality. Within each pair, we solely manipulated the edge cost randomness: Q1 and Q3 feature completely random costs (unpredictable branches), while Q2 and Q4 feature linearly predictable costs. 

As shown in Table~\ref{tab:do_505}, treatment datasets with highly predictable patterns (Q2, Q4) paradoxically execute \textit{slower} than their random counterparts (Q1, Q3) across all three platforms. CTMA reveals that highly accurate branch predictions, combined with deep pointer-chasing, act as ``Over-Speculation Congestion.'' The aggressive Frontend unleashes a Distributed Denial of Service (DDoS) attack on the Line Fill Buffers (LFB), severely exacerbating Backend memory bottlenecks. For example, on the Intel platform, the FB Full ratio (LFB congestion) surges from 42.5\% (Q1) to 59.2\% (Q2), and astonishingly from 6.4\% (Q3) to a complete deadlock at 100\% (Q4). Correspondingly, the Backend Bound metric increases from 28.2\% (Q1) to 30.0\% (Q2), and from 17.2\% (Q3) to 32.7\% (Q4).

\begin{table}[h]
\centering
\caption{Microarchitectural Profiles Across Dataset Treatments (\textit{mcf\_r})}
\resizebox{\linewidth}{!}{
\begin{tabular}{lcccccc}
    \toprule
    \textbf{CPU} & \textbf{Dataset} & \textbf{Br MPKI} & \textbf{L1D MPKI} & \textbf{LLC MPKI} & \textbf{IPC} & \textbf{Time (s)} \\ \midrule
    \multirow{4}{*}{\textbf{Intel}}  & \textbf{Q1}  & 24.58 & 84.18 & 43.16 & 0.26 & 777.7 \\
                                     & \textbf{Q2} & 20.42 & 81.96 & 40.75 & 0.28 & 859.5 \\ 
                                     & \textbf{Q3}  & 30.64 & 59.24 & 28.08 & 0.31 & 103.7 \\
                                     & \textbf{Q4} & 18.29 & 82.71 & 43.54 & 0.25 & 197.8 \\\midrule
    \multirow{4}{*}{\textbf{ARM}}    & \textbf{Q1} & 17.81 & 77.55 & 40.75 & 0.15 & 1284.5 \\
                                     & \textbf{Q2} & 15.69 & 75.33 & 39.74 & 0.15 & 1497.5 \\ 
                                     & \textbf{Q3} & 21.50 & 51.80 & 26.45 & 0.21 & 143.1 \\
                                     & \textbf{Q4} & 15.10 & 73.85 & 45.42 & 0.13 & 373.0 \\\midrule
    \multirow{4}{*}{\textbf{AMD}}    & \textbf{Q1} & 15.05 & 93.44 & 0.34  & 0.55 & 324.7 \\
                                     & \textbf{Q2} & 13.86 & 91.38 & 0.39  & 0.51 & 399.6 \\ 
                                     & \textbf{Q3} & 17.84 & 64.29 & 0.35  & 0.64 & 43.2 \\
                                     & \textbf{Q4} & 14.09 & 93.68 & 0.46  & 0.35 & 113.7 \\\bottomrule
\end{tabular}
}
\label{tab:do_505}
\end{table}


\textbf{Hardware Probe Cross-Validation:} To cross-validate this congestion, we apply the \textbf{Manual Speculation Throttling} probe. By compiling \texttt{\_\_asm\_\_ \_\_volatile\_\_("nop");} loops inside the hot function \texttt{price\_out\_impl()}, we establish a ``Hardware Traffic Light.'' Specifically, these injected \texttt{NOP} instructions force the pipeline to consume fetch and decode cycles without generating any actual data-memory requests. This acts like a red light, intentionally throttling the Frontend's commit rate and restricting the volume of speculative memory operations dispatched to the Backend. 

As shown in Table~\ref{tab:mcf-nop}, despite forcing the CPU to execute \textit{more} useless instructions, the elapsed time either remains stable or slightly \textit{decreases} across all three platforms. This proves that artificially slowing down the Frontend relieves Backend memory congestion, advocating for \textbf{dynamic speculation throttling mechanisms} in future CPU designs. A native hardware implementation could provide similar memory congestion relief without the instruction-delivery and retirement overhead of \texttt{NOP}s, enabling higher execution efficiency.

\begin{table}[h]
\centering
\caption{NOP Throttling Performance Paradox (\textit{mcf\_r})}
\label{tab:mcf-nop}
\resizebox{\linewidth}{!}{
\begin{tabular}{l|cc|cc|cc}
\toprule
\multirow{2}{*}{\textbf{Dataset}} & \multicolumn{2}{c|}{\textbf{Intel Xeon (s)}} & \multicolumn{2}{c|}{\textbf{ARM Kunpeng (s)}} & \multicolumn{2}{c}{\textbf{AMD EPYC (s)}} \\ \cline{2-7} 
 & Control & Treat (NOP) & Control & Treat (NOP) & Control & Treat (NOP) \\ \midrule
\texttt{ref}  & 794.99 & \textit{\textbf{789.97}} & 723.39 & \textit{\textbf{722.28}} & 261.22 & \textit{\textbf{260.50}} \\
\textbf{Q1}   & 457.60 & \textit{\textbf{455.30}} & \textit{\textbf{406.44}} & 406.57 & \textit{\textbf{138.35}} & 138.79 \\
\textbf{Q2}   & 502.51 & \textit{\textbf{500.48}} & 456.66 & \textit{\textbf{456.31}} & 155.84 & \textit{\textbf{154.24}} \\
\textbf{Q3}   & 71.59  & \textit{\textbf{70.92}} & 51.84  & \textit{\textbf{51.69}} & \textit{\textbf{21.22}} & 21.48 \\
\textbf{Q4}   & 101.07 & \textit{\textbf{100.91}} & 96.42  & \textit{\textbf{96.41}} & 35.89  & \textit{\textbf{35.82}} \\ \bottomrule
\end{tabular}
}
\end{table}

\subsection{Case Study: The Microarchitectural Ghost State and Pipeline Balancing on the Intel Xeon Gold 5120T}
Our causal treatments demonstrate that on the Intel Xeon Gold 5120T, modern CPU benchmarking workloads frequently trigger severe Frontend and Backend imbalances (e.g., LFB congestion in \textit{mcf\_r}, DSB reliance in \textit{xalancbmk\_r}, Branch MPKI surges \textit{cactuBSSN\_r}). During our SMT hardware probing, we observed a microarchitectural artifact that practically demonstrates the potential benefits of dynamically balancing these pipeline bottlenecks. Based on this artifact, we present a case study on microarchitectural bottleneck mitigation.

\textbf{Experimental Setup \& Observations:} We evaluated the SPEC CPU2017 and CPU2026 suites under three strictly controlled isolation states on the Intel Xeon gold 5120T:
\begin{itemize}
    \item \textbf{Clean-ST A1 (Baseline):} The workload runs isolated on a single pinned core (e.g., core 1).
    \item \textbf{Throttled-SMT (Hardware Throttling):} The workload process runs on core 1, while a constrained \texttt{\_mm\_pause()} process runs continuously on its sibling logical core (e.g., core 29 on our Intel Xeon Machine). This state structurally partitions the Backend queues (ROB/LSQ) and introduces coarse-grained bandwidth throttling.
    \item \textbf{Ghost-ST (The Artifact):} The constrained \texttt{\_mm\_pause()} process is forcefully terminated via the OS command \texttt{kill -9}. Immediately afterward, we re-execute the workload on core 1 while its SMT sibling (core 29) is reported as idle by the OS scheduler.
\end{itemize}

\begin{figure}[t]
    \centering
    \includegraphics[width=\linewidth]{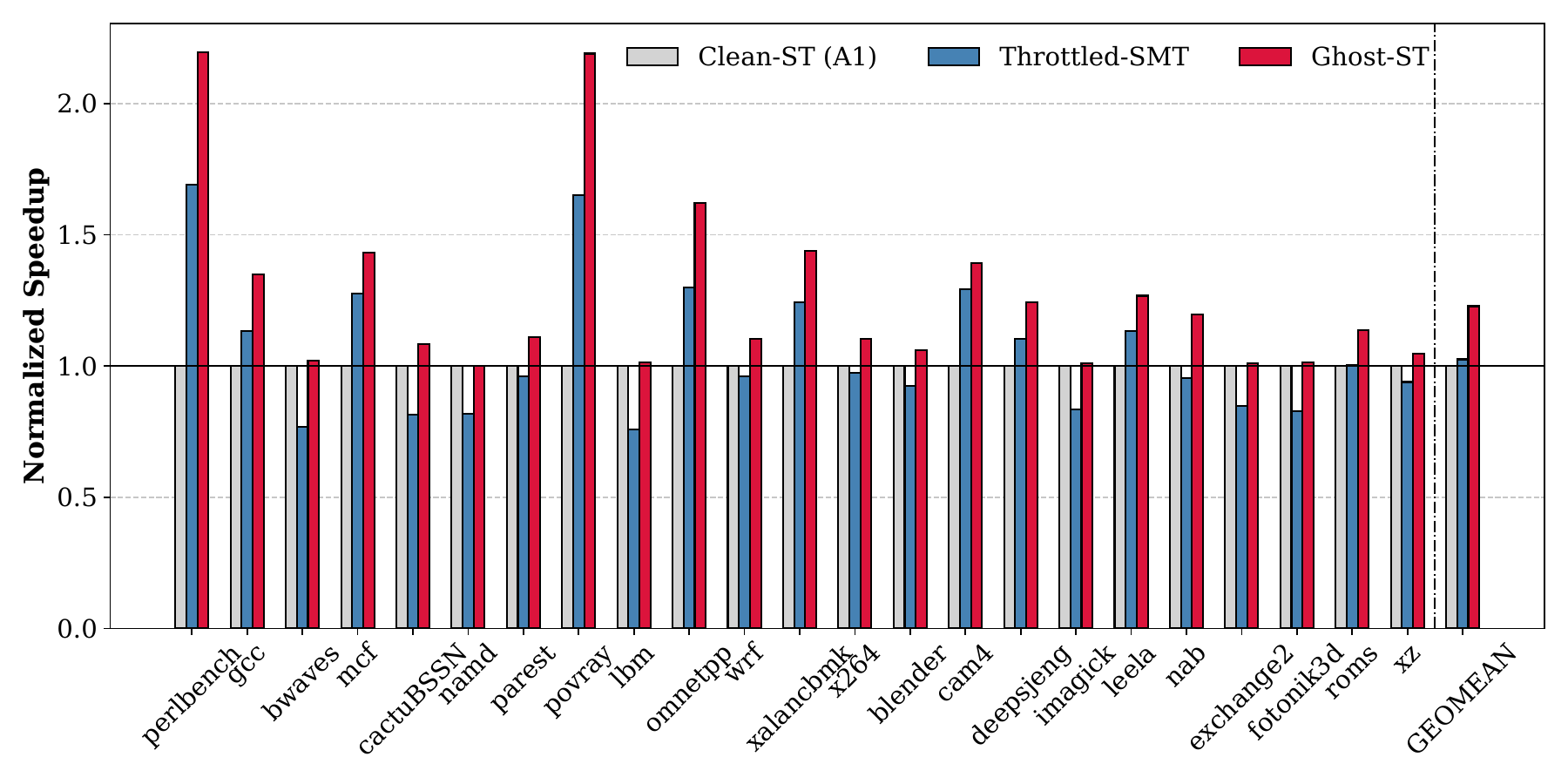}
    \caption{The Ghost State Speedup on the Intel Xeon Gold 5120T: Executing workloads in the stale pipeline (Ghost-ST) yields an average of 22.75\% speedup on SPEC CPU2017.}
    \label{fig:ghost_state_cpu2017}
\end{figure}

\begin{figure}[t]
    \centering
    \includegraphics[width=\linewidth]{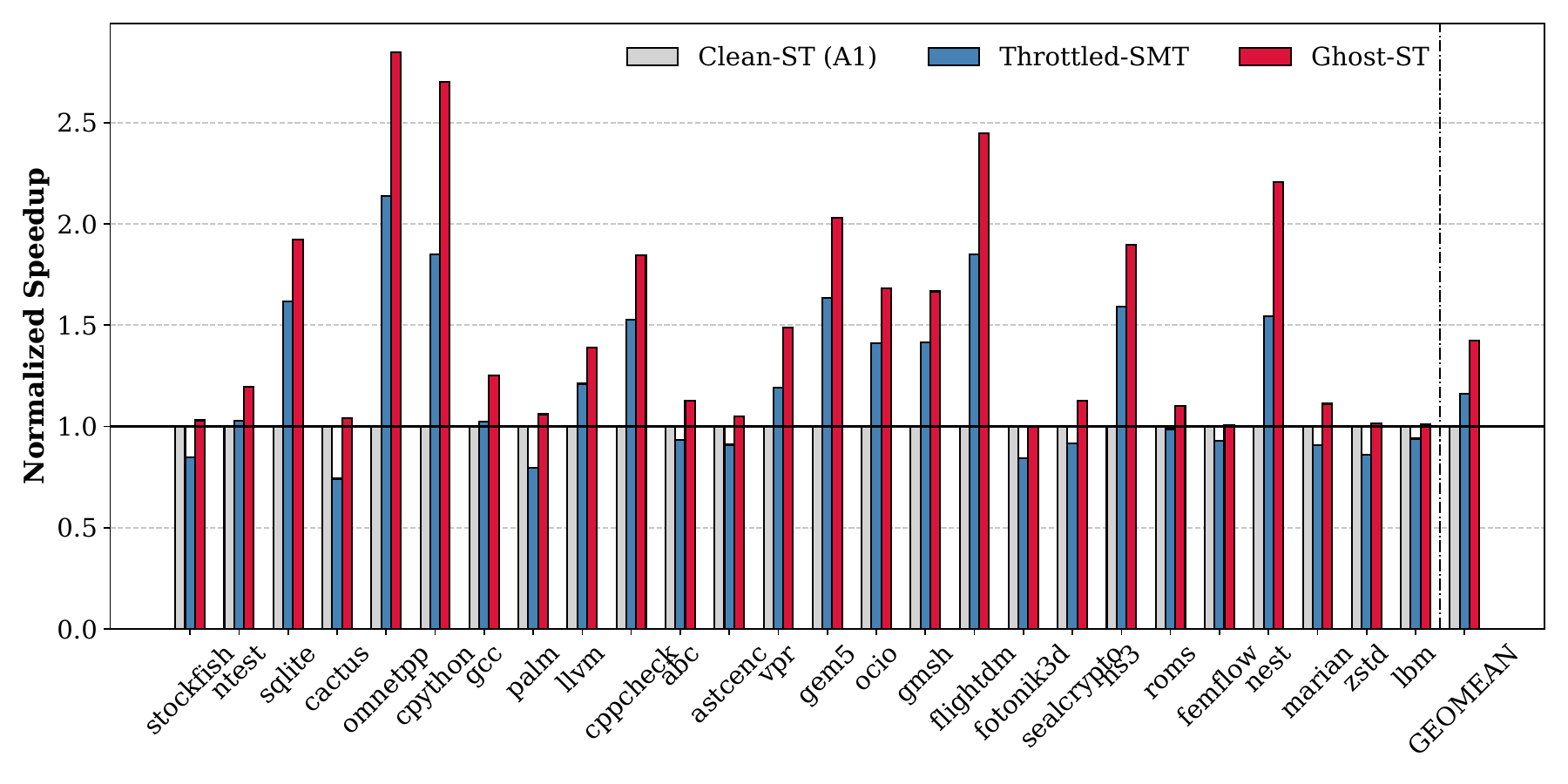}
    \caption{The Ghost State Speedup yields 42.20\% speedup on SPEC CPU2026.}
    \label{fig:ghost_state_cpu2026}
\end{figure}

As shown in Figure~\ref{fig:ghost_state_cpu2017} and Figure~\ref{fig:ghost_state_cpu2026}, even under the coarse-grained resource restriction of \textbf{Throttled-SMT}, workloads exhibit an average speedup of 2.52\% on SPEC CPU2017 and 16.09\% on SPEC CPU2026 compared to Clean-ST. This confirms that for modern intensive benchmarks, blindly allowing a wide pipeline to operate at maximum capacity often degrades performance due to congestion; restricting the pipeline bandwidth can actually yield positive returns, with higher pressure workloads (CPU2026) benefiting more significantly.

However, the most notable anomaly occurs in the \textbf{Ghost-ST}. After the \texttt{kill -9} termination, the OS clears the process context; yet, the hardware fails to immediately reset dynamic pipeline configurations, leaving a microarchitectural "Ghost State".
Executing workloads in this state yields an average speedup of 22.75\% on SPEC CPU2017 (ranging from 0.09\% to 119.45\%) and 42.20\% on CPU2026 (ranging from 0.19\% to 184.73\%) over the Clean-ST baseline.






\begin{figure}[t]
    \centering
    \includegraphics[width=\linewidth]{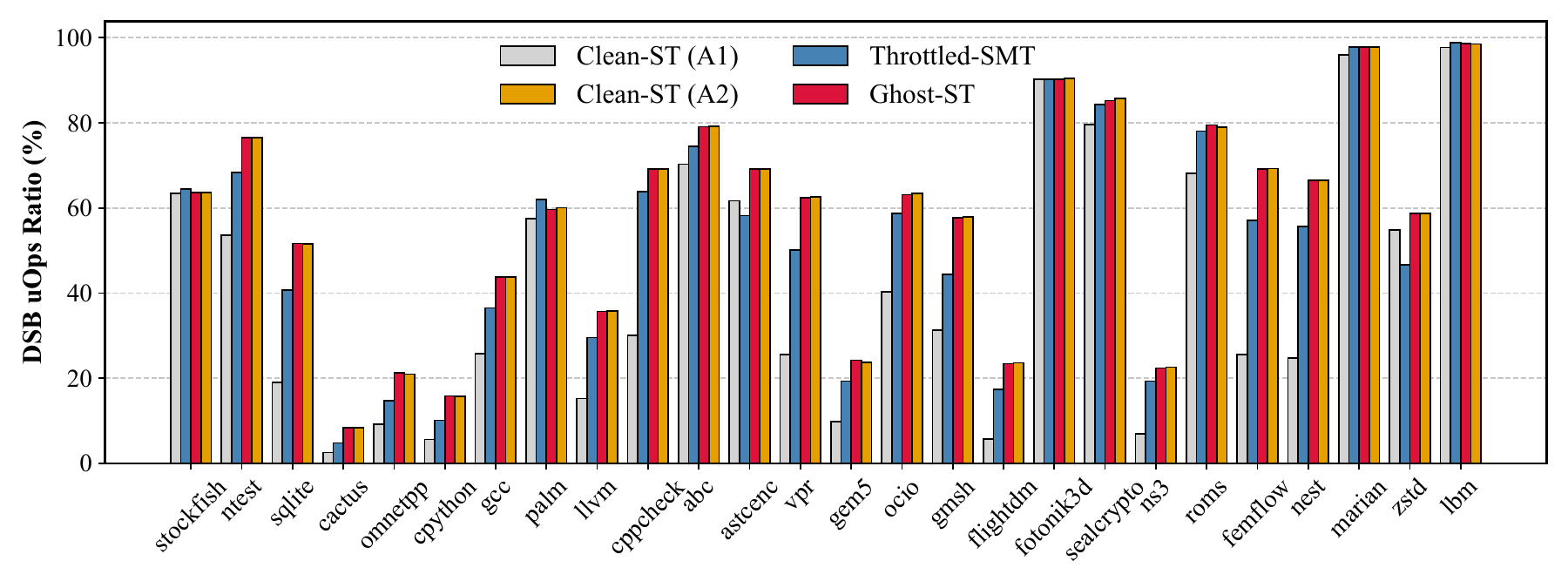}
    \caption{DSB uOps Ratio surges in  Ghost-ST on SPEC CPU2026, matching the unmitigated state.}
    \label{fig:cpu2026_dsb_uops_ratio}
\end{figure}
\begin{figure}[t]
    \centering
    \includegraphics[width=\linewidth]{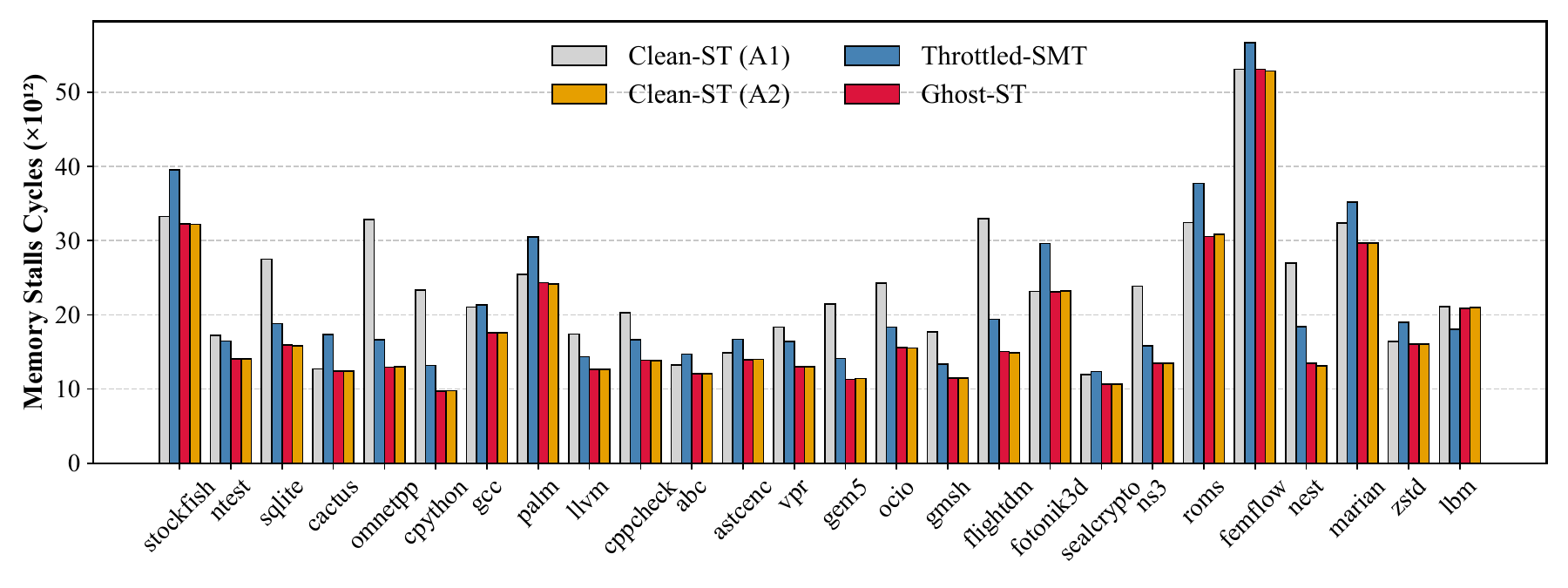}
    \caption{Memory Stall Cycles drastically decrease in  Ghost-ST on SPEC CPU2026.}
    \label{fig:cpu2026_mem_stall_cycle}
\end{figure}




\textbf{Root Cause Analysis:} Through CTMA profiling on SPEC CPU2026 (Figures~\ref{fig:cpu2026_dsb_uops_ratio} to \ref{fig:cpu2026_mem_stall_cycle}), we observed that Ghost-ST exhibits a massive surge in the DSB uOps ratio alongside a drastic reduction in Memory Stall Cycles (The profiling results on SPEC CPU2017 are similar to SPEC CPU2026). Crucially, the microarchitectural signature of Ghost-ST exactly matches a clean baseline core booted with the kernel parameter \texttt{GRUB\_CMDLINE\_LINUX\_DEFAULT="spectre\_v2=off"} (denoted as \textbf{Clean-ST A2}). 
This evidence implies that the abrupt process termination bypasses the OS kernel's security context-scrubbing routines, inadvertently disabling the severe DSB suppression imposed by Spectre V2 mitigations, thereby unleashing the Frontend.



\textbf{Microarchitectural Implications: Dynamic Speculation Throttling.} 
Observed on a typical Intel Xeon Gold 5120T processor, this ``Ghost State'' paradox illuminates a profound design flaw: modern wide-issue pipelines are often substantially unbalanced. When security mitigations arbitrarily throttle the Frontend (as seen in Clean-ST), performance suffers. Conversely, when unrestricted speculation creates Backend memory traffic jams (as seen in \textit{mcf\_r}), static hardware queues fail, and artificially throttling the bandwidth (as seen in Throttled-SMT) actually yields positive effects.

Rather than blindly expanding pipeline buffers (which hit marginal utility ceilings) or applying rigid, one-size-fits-all security throttles, future CPU designs should incorporate \textbf{dynamic speculation throttling mechanisms}. Processors require a closed-loop monitoring interface that observes Backend congestion (e.g., LFB occupancy limits) to dynamically throttle or unleash Frontend speculation. CTMA demonstrates that strategically balancing the pipeline flow yields far greater dividends than brute-force capacity scaling.

\section{Conclusion and Future Work}
\label{sec:conclusion}

Traditional Top-Down Microarchitecture Analysis (TMA) suffers from \textit{causal blindness}. By profiling static workloads, it blindly attributes performance variations solely to microarchitectural efficiency (CPI), fundamentally ignoring algorithmic instructions bloat and dynamic bottleneck migrations. 

To cure this causal blindness, we proposed \textbf{Causal TMA (CTMA)}. Grounded in the object causal model (OCM), CTMA pairs  software treatments with  hardware probe (e.g., core frequency clamping, SMT resource partitioning) cross-validation across Intel, AMD, and ARM platforms. We systematically proved that predictable data triggers ``Over-Speculation Congestion,'' geometric distortions shift pipelines from capacity-bound to fill-rate-bound, and Op-cache footprints dictate SMT overlap. Furthermore, uncovering an average of 42.20\% performance boost of the ``Ghost State'' artifact exposed the severe penalties of unbalanced pipelines. Ultimately, CTMA prevents architects from falling into correlational traps and optimizing the wrong bottlenecks.

CTMA lays the groundwork for a new performance analysis and CPU DSE paradigm. We plan to extend this work in below directions:
1) \textbf{Simulator Integration:} Extending our hardware probing interface to simulators for register-level causal interventions beyond the limits of commercial silicon. 2) \textbf{Dynamic Speculation Throttling:} 
Developing closed-loop hardware controllers that dynamically regulate Frontend dispatch based on Backend congestion, achieving pipeline balance without expanding pipeline buffer.



\bibliographystyle{IEEEtranS}
\bibliography{refs}


\end{document}